\documentclass[number,3p,times,preprint,review,fleqn]{elsarticle}
\usepackage{comment}
\usepackage{verbatim}

\usepackage{graphicx}
\usepackage[figuresright]{rotating}
\usepackage[percent]{overpic} % use text on graphics
\usepackage{caption} % for using ContinuedFloat

\usepackage{makecell}
\usepackage{tabularx} % complex tabular
\usepackage{booktabs} % complex tabular
\usepackage{multirow} % 
\usepackage{threeparttable} % for note in the table
\usepackage{stackengine}
\newcommand\xrowht[2][0]{\addstackgap[.5\dimexpr#2\relax]{\vphantom{#1}}}
\usepackage{amsmath,natbib,mathrsfs,stmaryrd,titletoc,subfigure}
\usepackage{amssymb,amsfonts,amsthm,stmaryrd}
\usepackage{upgreek}
\usepackage[ruled,vlined]{algorithm2e}

\usepackage[textsize=tiny]{todonotes} % things to be done

\usepackage[section]{placeins}
\usepackage[pdftex,
CJKbookmarks=false, bookmarksnumbered=true, bookmarksopen=true,
colorlinks=true,
citecolor=blue, linkcolor=red, anchorcolor=green,
urlcolor=blue, pagebackref=true]{hyperref}
\hypersetup{pdfauthor={Name}}
\usepackage[capitalise]{cleveref} % must be loaded after 'hyperref'

\usepackage{color}

\newcommand{\bfx}{\boldsymbol{x}}
\newcommand{\bfJ}{\boldsymbol{J}}
\newcommand{\bfu}{\boldsymbol{u}}
\newcommand{\bfn}{\boldsymbol{n}}
\newcommand{\bfsigma}{\boldsymbol{\sigma}}

\newcommand{\bfepsilon}{\boldsymbol{\epsilon}}

\graphicspath{{./figures/}}

\numberwithin{equation}{section}

\newcounter{remark}[section]
\newcommand{\remark}{\noindent \textbf{Remark} \refstepcounter{remark} \textbf{\theremark} \;}

\SetAlCapSkip{1em} %=> vertical space between algorithms and text

\crefname{figure}{Figure}{Figure}
\crefname{equation}{Eq.}{Eqs.}
\definecolor{darkgray}{rgb}{0.95,0.95,0.95}

\usepackage[most]{tcolorbox}

\crefname{boxcref}{Box}{Boxes}

\newtcolorbox[use counter=boxcref]{mybox}[2]{
	enhanced,
	width = 13cm,
	center,
	center title,
    halign=center,
	valign=center,
	label type=boxcref,
	label = #1,
	colback=gray!5!white,
	colframe=gray!75!black,
	title = Box~\thetcbcounter:~#2
}

\newtcolorbox[use counter=boxcref]{mybox2}[2]{
	enhanced,
	width = 16cm,
	center,
	center title,
    halign=center,
	valign=center,
	label type=boxcref,
	label = #1,
	colback=gray!5!white,
	colframe=gray!75!black,
	title = Box~\thetcbcounter:~#2
}

\journal{Elsevier}
\begin{document}
	
	\begin{frontmatter}
		
		
		% use the tnoteref command within \title for footnotes;
		% use the tnotetext command for theassociated footnote;
		% use the fnref command within \author or \address for footnotes;
		% use the fntext command for theassociated footnote;
		% use the corref command within \author for corresponding author footnotes;
		% use the cortext command for theassociated footnote;
		% \author{Name\corref{cor1}\fnref{label2}}
		% \address{Address\fnref{label3}}
				
		% \title{A length-scales insensitive cohesive phase-field model for chemo-mechanical inter-/trans-granular fractures in cathode particles}

		\title{A length-scale insensitive cohesive phase-field interface model: application to concurrent bulk and interface fracture simulation in Lithium-ion battery materials}

		% \title{A cohesive phase-field diffusive interface model with energy equivalence to sharp crack: application to inter-/trans-granular fractures in cathode particles}

		% \title{A cohesive phase-field diffusive interface model with energy equivalence to sharp representation: application to inter-/trans- granular fractures in cathode particles}
		% image-based reconstructed 3D NMC cathode particles}
		
		%\tnotetext[label1]{Dedicate to }
		
		% *author information
		% optional format:

		% authors from TUD-MFM
		\author[label1]{Wan-Xin Chen\corref{cor1}}
		\address[label1]{Mechanics of Functional Materials Division, Institute of Materials Science, Technischen Universität Darmstadt, Darmstadt 64287, Germany}
		\cortext[cor1]{Corresponding author(s).}
		\ead{wanxin.chen@tu-darmstadt.de}
		%
		% \author[label2]{Luis J. Carrillo\corref{cor1}}
		%
		\author[label1]{Xiang-Long Peng}
		
		\author[label5]{Jian-Ying Wu}
		\address[label5]{State Key Laboratory of Subtropical Building and Urban Science, South China University of Technology, Guangzhou 510641, China}

		%\author[label2]{Jeffery M. Allen}
		%\address[label2]{Computational Science Center, National Renewable Energy Laboratory, Golden, CO 80401, USA} 
		
		\author[label3]{Orkun Furat}
		\address[label3]{Institute of Stochastics, Ulm University, 89069 Ulm, Germany}
		
		\author[label3]{Volker Schmidt} 
		
		%\author[label4]{Kandler Smith}
		%\address[label4]{Center for Energy Conversion and Storage Systems, National Renewable Energy Laboratory, Golden, CO 80401, USA}
		%
		% \author[label2]{Sarbajit Banerjee\corref{cor2}}
		% \ead{banerjee@chem.tamu.edu}
		%
		\author[label1]{Bai-Xiang Xu\corref{cor1}}
		\ead{xu@mfm.tu-darmstadt.de}
		%
		% \address[label1]{Mechanics of Functional Materials Division, Institute of Materials Science, Technischen Universität Darmstadt, Darmstadt 64287, Germany}
		% \address[label2]{Department of Chemistry, Texas A$\&$M University, TX 77843-3255, USA}
		% %\address[label2]{Computational Science Center, National Renewable Energy Laboratory, Golden, CO 80401, USA}	
		% \cortext[cor1]{These authors contributed equally to this work.}
		% \cortext[cor2]{Corresponding author(s).}
		
		%\def\thefootnote{*}\footnotetext{These authors contributed equally to this work.}\def\thefootnote{\arabic{footnote}}

		%\author[label3]{Shahed Rezaei}
		%\address[label3]{Access e.V., Aachen, Germany}
		
		%\author[label5]{\\Avtar Singh}
		%\address[label5]{Center for Energy Conversion and Storage Systems, National Renewable Energy Laboratory, Golden, CO 80401, USA}
		
		% \author[label2,label1]{Bai-Xiang Xu\corref{cor1}}
		%\author[label1]{Bai-Xiang Xu\corref{cor1}}
		%\cortext[cor2]{Corresponding author.}
		%\ead{xu@mfm.tu-darmstadt.de}
		
		%\maketitle
		%\tableofcontents
		
		% *abstract
		\begin{abstract}
			
		A new cohesive phase-field (CPF) interface fracture model is proposed on the basis of the Euler-Lagrange equation of the phase-field theory and the interface fracture energy check w.r.t. that of the cohesive zone model. 
		It employs an exponential function for the interpolation of fracture energy between the bulk phase and the interface, while the effective interface fracture energy $\tilde{G}_i$ is derived in such a way that the integrated phase-field fracture energy across the diffusive interface region remains consistent with the sharp interface fracture energy $G_i$ defined in the classical cohesive zone model. This consistency is the key to ensure that the numerical results remain insensitive to the choice of length-scale parameters, particularly the regularized interface thickness $L$ and the regularized fracture surface thickness $b$. By employing this energy consistency check, various CPF interface models in the literature are reviewed. Besides the length-scale insensitivity, the proposed CPF interface model offers further advantages. Thanks to the fact that the exponential interpolation function can be obtained conveniently from the relaxation solution of an Allen-Cahn equation, the proposed CPF model is advantageous over other models with high flexibility in handling structures containing complicated interface topology. In order to demonstrate this merit and to check the length-scale insensitivity in multiphysics context, the proposed CPF interface model is employed further to derive a thermodynamically consistent chemo-mechanical model relevant to Lithium-ion battery materials. Finite element simulation results of the concurrent bulk and interface fracture in polycrystalline electrode particles, reconstructed from images with segmented interfaces, confirm the expected computational advantages and the length-scale insensitivity in chemo-mechanical context. 

		% 
		
		%To capture cracks happening at the GBs within the phase-field framework, we introduced a general methodology diffusively representing the interface, incorporating an elaborately determined effective interfacial fracture toughness, which ensures the fracture energy equivalent to a sharp interfacial crack.
		% 
		%Furthermore, this approach is extended to multi-dimensional scenarios and seamlessly integrated into the coupled chemo-mechanical cohesive phase-field fracture model in a thermodynamically consistent manner.
		% 
		%Through representative benchmark examples, we validate the capability of our proposed model to accurately simulate inter-granular fracture behaviors of cathode particles. Our numerical results closely align with predictions from the cohesive zone model, and highlight the advantageous trait of being insensitive to variations in both phase-field and interface length-scale parameters. Finally, the proposed model is applied to image-based reconstructed 3D polycrystalline cathode particles. The resulting findings offer predictive and valuable insights into key fracture mechanisms for mechanical degradations within NMC particles.
			
		\end{abstract}
		
		% *abstract
		\begin{keyword}
			Cohesive phase-field fracture; Cohesive zone model; Interface fracture energy check; Length-scale insensitivity; Chemo-mechanical fracture; Concurrent bulk and interface fracture
			
			
			% MSC codes here, in the form: \MSC code \sep code
			% or \MSC[2008] code \sep code (2000 is the default)
		\end{keyword}
		
	\end{frontmatter}
	
	% main text***************************************************************************************
	%%%%%%%%%%%%%%%%%%%%%%
	\section{Introduction}
	\label{sec:Introduction}
	
	% \begin{itemize}
	% 	\item fracture of lithium ion batteries, interface and bulk fracture, cohesive zone model to phase-field approach
	% 	\item phase-field fracture of lithium-ion batteries
	% 	\item phase-field fracture of bulk/interface fracture
	% \end{itemize}

    Advanced material systems are inherently heterogeneous, often comprising multiple phases or distinct bulk-interface regions in their complex microstructures. Examples of such materials include engineering materials like fiber-reinforced cement composites \cite{Li2018CA}, coating systems \cite{thermal-barrier-coatings}, multiphase composites \cite{Zhang2022CMAME}, and composite laminates \cite{composite-laminates}, as well as biomaterials like bones \cite{Sandino2014JB} and energy materials like Lithium-ion batteries (LIBs)\cite{Ryu2018CM}. A common phenomenon in these materials is the concurrent occurrence of bulk and interface fracture, which degrades the overall capacity and performance of the materials and structures \cite{capacity_fading}. This necessitates the development of computational models to accurately recapture and predict these fracture behaviors, thereby enhancing the reliability and durability of these advanced materials.

    To model the cracking behaviors at interfaces, the cohesive zone model (CZM) \cite{CZM1,CZM2} is widely used due to its ease of implementation for interfacial properties and its accuracy in reproducing experimental results. In the CZM, the discontinuous displacement jump at the interface is explicitly represented with cohesive interface element \cite{CZM2002}, and a constitutive relation between the (local) traction vector $\tilde{\boldsymbol{t}}$ and the (local) displacement jump vector $[\![ \tilde{\boldsymbol{u}} ]\!]$ is formulated, which is known as the traction-separation law (TSL). However, this approach exhibits challenges in modeling arbitrary cracks propagation within the bulk phase \cite{CZM2002}. In the last two decades, the phase-field fracture model originally proposed by \cite{AMM2009JMPS,Miehe2010IJNME}, which utilizes a damage-like scalar variable to regularize the sharp crack and minimizes the total system energy within a variational framework, has been a popular candidate in modeling complex evolution of arbitrary cracks in the solids. In particular, \cite{Wu2017JMPS,Wu2018JMPS} proposed a cohesive phase-field (CPF) fracture model utilizing the parameterized energy degradation function and crack geometry function, which can reproduce the general softening laws of the CZM and ensure the results are insensitive to the choice of internal phase-field length-scale parameter $b$. Despite its extensive applications in modeling fracture problems in homogeneous materials, the phase-field model's ability to accurately simulate arbitrary cracks propagation in heterogeneous systems, which include both bulk phases and interfaces, remains limited.

    The cohesive zone model for interface fractures and the phase-field model for bulk phase cracks, which represent discrete and smeared crack configurations respectively, are technically two distinct approaches, and integrating them into a unified framework is challenging. However, bulk and interface fractures can happen simultaneously in heterogeneous systems, which are regulated by the energy-based material physics. In this context, we anticipate to utilize an unified material model in this manuscript, which can recapture both types of fracture at the same time. As reviewed following in \cref{sec:phase-field-interface}, the phase-field fracture model is promising to be extended to recapture the interface fracture, while it still remains challenging and awaits for further improvements regarding the energy equivalency between the phase-field framework and the CZM, as well as the length-scale sensitivity issue of phase-field model, which will be the main focus of this paper. Moreover, the concurrent bulk and interface fracture in engineering materials happen mostly under multi-physical circumstance, as can be seen in \cref{sec:chemo-mechanical}, an important prototyping example is chemo-mechanical inter- and trans-granular fractures in Lithium-ion battery cathode materials. Extending the unified model to address chemo-mechanically coupled bulk and interface fracture and verifying its robustness in a multi-field context are also challenging points to be investigated in this paper.

\subsection{Phase-field modeling of interface fracture}\label{sec:phase-field-interface}
	
		In order to model and simulate crack propagation in a heterogeneous system including bulk phase and interface, several attempts in recent years have been made to develop the phase-field model, trying to take into account the influence of interface on the fracture evolution process. One approach integrates the phase-field method for bulk fracture with the cohesive zone model for interface crack, wherein the displacement jump across the interface is represented using either cohesive interface element \cite{Paggi2017CMAME,Paggi2017CS} or a regularized form based on the phase-field approximation \cite{Nguyen2016CMAME}. This approach presents challenges in defining the coupling relations between the phase-field and cohesive zone damage, and it is cumbersome to numerically implement two types of material constitutive laws. The other popular approach is to represent the interface in a diffused manner within the phase-field framework, namely, the fracture energy in the domain is interpolated between that of the bulk phase and the interface \cite{Nguyen2017CM,Zhang2020CS,Li2020JMPS}, which can be conveniently implemented based on an auxiliary indicator solved from the Allen-Cahn equation \cite{Nguyen2016CMAME}. This diffusive approach provides significant flexibility in modeling crack propagation at interfaces, which has inspired extensive applications \cite{Li2021CMAME,Nguyen2019JCP}, including Lithium-ion batteries \cite{Avtar2022JMPS,Emilio2023JMCA,Avtar2024EML} in the multi-physical context. However, as will be shown later in this paper, the aforementioned diffusive treatment on the interface fails to ensure the (integrated) fracture energy within the phase-field framework consistent with that defined in the cohesive zone model, which correspondingly leads to the length-scale sensitivity \cite{Wu2018JMPS}, i.e., parameters regularizing the crack surface $b$ and interface $L$ have essential impacts on fracture and mechanical behaviors of structures.

		To address the aforementioned length-scale sensitivity issue, \cite{Hansen2019CMAME, Hansen2020EFM} first proposed to use the effective interface fracture energy w.r.t the choices of length-scale parameters ($b$ and $L$) in the phase-field interface model. This approach, which is referred to as Model-S1 in \cref{sec:model-S1}, tries to constrain the (integrated) fracture energy dissipation to be equivalent to that defined in the cohesive zone model for a sharp interface. Nevertheless, these studies have found that interface fracture resistance is always underestimated in the simulations, which is attributed to the use of the phase-field profile of a homogeneous bulk phase when solving the energy consistency constraint equation. On the basis of that, \cite{Yoshioka2021CMAME, Zhou2021IJSS} introduced the Euler-Lagrange equation to analytically express the phase-field profile in a heterogeneous domain with both bulk and diffusive interface (see Model-S2 in \cref{sec:model-S2}). They successfully determined the proper effective fracture energy to achieve interface fracture energy consistency and interface-width insensitivity. It's worth noting that, above derivations of the effective interface fracture energy are all applicable to the stair-wise interpolation of materials properties between the bulk phase and the interface, thanks to the phase-field profile being analytically solvable from the Euler-Lagrange equation. However, as proved by extensive applications under purely mechanical \cite{Nguyen2017CM,Zhang2020CS,Li2020JMPS} and multi-physical contexts like Lithium-ion batteries \cite{Avtar2022JMPS,Emilio2023JMCA,Avtar2024EML,Chen2024JPS}, the exponential interpolation of material property implemented with an auxiliary Allen-Cahn indicator offers greater flexibility compared to the stair-wise one in handling structures containing complex interface topology, particularly for the image-based reconstructed 3D polycrystalline microstructure \cite{Chen2024JPS}; nevertheless, such length-scale insensitive phase-field interface model still remains absent in the literature.
  
	\subsection{Chemo-mechanical inter- and trans-granular fractures in LIB cathode materials} \label{sec:chemo-mechanical}
		Lithium-ion batteries have emerged as a revolutionary technology, powering a diverse range of essential devices from smartphones to electric vehicles, and serving as one of the promising candidates for next-generation energy storage technology \cite{Battery2018,Battery2021}. Despite their promising merits and widespread commercial deployment, LIBs suffer from significant performance and capacity fading in the cycles of charging and discharging, in particular, damage and fracture induced chemo-mechanical degradation \cite{Zhao2019JPSReview} has been recognized as one of the key mechanisms influencing their longevity and reliability. Extensive experimental research is underway to elucidate the intensive existence of cracks and their critical role in causing deterioration in battery systems, e.g., lithium dendrite formation \cite{dendrite1} induced crack propagation in the electrolyte, fracture in the silicon anode particle due to significant volume change\cite{silicon1}, fracture of cathode particles \cite{Xu2017JES}, failure of solid electrolyte \cite{electrolyte}, interfacial delamination \cite{delamination}, and so on. Those fractures within the LIBs impact the electrochemically active surface area, hinder charge-transfer reactions at the electrode/electrolyte interfaces, and eventually lead to increased cell resistance and reduced available capacity \cite{delamination,capacity_fading}.

		In this paper, we will focus on the damage and fracture behaviors of cathode particles in the LIBs. Most commonly available cathodes, such as LiCoO$_2$, LiMn$_2$O$_4$ and LiNi$_x$Mn$_y$Co$_{1-x-y}$O$_{2}$ (NMC) and so on, exhibit a polycrystalline microstructure \cite{Furat2021JPS} in nature, those secondary particles (order 10-20 $\upmu$m) are comprised of randomly oriented single grains, also referred to as primary particles \cite{grain}. The interfaces where crystalline grains meet are termed as grain boundaries (GBs). During (dis)charging cycle, Li (de)intercalation can induce anisotropic volume change (expansion/shrinkage) \cite{Song2016PCCP}, which can generate inhomogeneous stress distribution in the structures. Typically, GBs with lower mechanical failure resistances \cite{capacity_fading,Nakajima2021CM} act as potentially starting points with damage nucleation and crack propagation, forming inter-granular (interface) fracture at the GBs \cite{capacity_fading,kim2015new}; in addition, with introduced defects (e.g., vacancy and surface curvature) and the random grains' orientations \cite{lin2014surface}, trans-granular (bulk) cracks within the primary particles can also be observed in the experiments \cite{yan2017intragranular,qian2020single}. Furthermore, in a multi-physically coupled context, cracks damage the pathways within the electrode structure for lithium transport, leading to degradation in the chemical diffusion process \cite{Klinsmann2015JES}, potential grain isolation \cite{Chen2024JPS}, and overall deterioration of cathode capacity \cite{Ryu2018CM}. 

		The aforementioned evidences and insights garnered from experiments provide a valuable opportunity for conducting numerical investigations into the chemo-mechanical behaviors of polycrystalline cathode particles. Several computational endeavors have been undertaken to model the chemo-mechanical behaviors in typical LIBs electrode, e.g., lithium (de)intercalation induced mechanical stresses \cite{villani2014fully,Tian2020JES,stein20143d,Leo2014JMPS}, phase segregation captured with a Cahn-Hilliard phase-field model \cite{Leo2014JMPS,Zhao2015CMAME}, electrochemical performance and embrittlement degradation over (dis)charging cycles (fatigue) \cite{Bucci2017JMCA,Emilio2022JPS,taghikhani2023electro}, influences of mechanical and transport features of grain boundaries \cite{Singh2020IJP,Bai2020SM,Shahed2021JMPS}, etc. See \cite{Zhao2019JPSReview,Leo2021M,McMeeking2023review} for comprehensive reviews on chemo-mechanical modeling of LIBs. 
  
        With a particular focus on damage and fracture behaviors in the LIBs cathode materials in this paper, here available chemo-mechanically coupled models are reviewed. Inspired by its widespread use in purely mechanical context, the cohesive zone model \cite{CZM2002} has been successfully extended to chemo-mechanical scenario and is likely the most popular candidate among other available options, see \cite{Sun2016EML,Bucci2017JMCA,Xu2018JMPS,Zhang2019IJMS,Shahed2021JMPS,Bai2021IJSS,Singh2020IJP,taghikhani2023electro} for numerous applications in LIBs. The CZM excels in recapturing cracks propagation along the GBs and the corresponding across-GBs degradation in electrode particles. However, its ability to predict arbitrary trans-granular cracks within the grains is limited. Comparatively, phase-field fracture model originally proposed by \cite{BFM2008,Miehe2010IJNME}, which 
        %describes the crack as the evolution of a (scalar) damage variable in the continuum context and 
        demonstrates advantages in simulating complex evolution of arbitrary cracks, showcases robust capabilities in modeling chemo-mechanical fracture and the induced degradation in the LIBs, see considerable successes in \cite{Klinsmann2015JES,Zuo2015PCCP,Xu2016GAMMMITT,Zhao2016CMAME,Miehe2016IJNME,Karma2019JMPS,Boyce2022JPS,Emilio2022JPS,Cao2023JAC,Zhang2024JMPS}. Despite the noteworthy contributions, those phase-field models for LIBs operate at the macro-scale level and overlook the intricate polycrystalline microstructure comprised of diverse grains and complex GBs, where the latter typically exhibits weaker fracture properties \cite{capacity_fading,Nakajima2021CM}. Consequently, diverse fracture modes including inter- and trans-granular cracks \cite{kim2015new,yan2017intragranular,qian2020single} observed in the experiments can not be recaptured in aforementioned simulations.

		To account for the heterogeneity within the cathode particle microstructure and predict both inter- and trans-granular failure modes, several endeavors have been made to incorporate interface fracture property into the phase-field framework, e.g. the sharp interface fracture energy is interpolated with that of the bulk phase in the domain, through either a stair-wise function \cite{Shahed2023TAFM,Chen2024JPS} (refer to Model-S0, see \cref{sec:model-S0}) or an exponential function \cite{Avtar2022JMPS,Emilio2023JMCA,Avtar2024EML} (refer to Model-E0, see \cref{sec:model-E0}) based on the predefined auxiliary indicator solved from the Allen-Cahn equation \cite{Nguyen2016CMAME}. Such approach can qualitatively simulate (and distinguish) diverse cracking modes at the GBs/interfaces or within the grains/bulks, whereas it will be shown later in this paper, it can not ensure the evaluated interface fracture energy (i.e., the amount of energy consumed for unit crack surface) within the phase-field framework to be consistent with that of a sharp interface defined in the cohesive zone model \cite{CZM2002}. Furthermore, the fracture energy dissipation for interface crack strongly depends on the choices of two length-scale parameters $b$ and $L$, which are for respectively regularizing the crack surface and interface, so that length-scale sensitive outcomes would be anticipated, e.g., different fracture patterns and quantitative structural responses (e.g., reaction force) are obtained with varying $b$ or $L$ in the simulations. $\Box$

	% ------------------------------------------------------------------------------
	In this paper, our primary goal is to develop a new length-scale insensitive cohesive phase-field interface fracture model and apply to the chemo-mechanical simulations of polyscrystalline structures with complicated GBs topology within the LIB materials. We utilized an exponential interpolation of fracture energy between bulk phase and interface on the basis of the solution of an Allen-Cahn equation. The effective interface fracture energy is introduced and derived based on the Euler-Lagrange equation of the phase-field theory, so that the integrated phase-field fracture energy can be ensured to be equivalent to the sharp value defined in the cohesive zone model. With multi-dimensional and multi-physical extensions in a thermodynamically consistent manner, the model can simulate the chemo-mechanical inter-/trans- granular fractures in the Lithium-ion battery materials, particularly with following merits expected, e.g., consistency with predictions from the CZM, length-scale insensitivity and flexibility dealing with complicated GBs topology.

	The remainder of this paper is organized as follows. \cref{sec:bulk} recaps the phase-field model for bulk fracture. In \cref{sec:interface}, we propose a new phase-field interface model and compare it with several existing models from the literature, in particular, we highlight the merit of the proposed model in ensuring that the interface fracture energy is consistent with that defined in the cohesive zone model. In \cref{sec:phase-field}, the proposed phase-field interface model is extended to chemo-mechanical scenario in a thermodynamically consistent manner. Governing equations, constitutive laws, and couplings among different physics are properly derived. \cref{results} presents several representative numerical examples, e.g., purely mechanical and chemo-mechanical benchmarks underscore the foundational advantages of the proposed model, subsequently, simulations of inter- and trans-granular fracture behaviors in image-based reconstructed 3D NMC polycrystalline particles are conducted. The relevant conclusions and outlooks are drawn in \cref{sec:conclusions}.
	
	\textit{Notation}. Compact tensor notation is used in the theoretical part of this paper. As general rules, scalars are denoted by italic light-face Greek or Latin letters (e.g., $a$ or $\lambda$); vectors, second- and fourth-order tensors are signified by italic boldface minuscule, majuscule and blackboard-bold majuscule characters like $\boldsymbol{a}$, $\boldsymbol{A}$ and $\mathbb{A}$, respectively. The inner products with single and double contractions are denoted by `$\cdot$' and `:', respectively. 
	
	% The dyadic products '$\otimes$' and '$\symotimes$' are defined as
	% %
	% \begin{align}
		% 	\big( \boldsymbol{a} \otimes \boldsymbol{B} \big) _{ijk} = a_i B_{jk} 
		% 	\qquad
		% 	\big( \boldsymbol{a} \symotimes \boldsymbol{B} \big) _{ijk} = \dfrac{1}{2} \big( a_j B_{ik} + a_k B_{ij} \big) \notag
		% \end{align}
	% 
	%\clearpage
	
	%%%%%%%%%%%%%%%%%%%%%%%%%%%%%%%%%%%%%%%%%%%%%%%%%%%%%%%%%%%%%%%%%%%%%%%%%%%%%%%%%%%%%%%

	% ***********************************************
	\section{Recap of phase-field bulk fracture model}
	\label{sec:bulk}

	In this section, the fracture energy dissipation due to bulk crack propagation within the phase-field framework is briefly recalled. 

	\begin{figure}[htbp]
		\centering
		\includegraphics[width=1.0\textwidth]{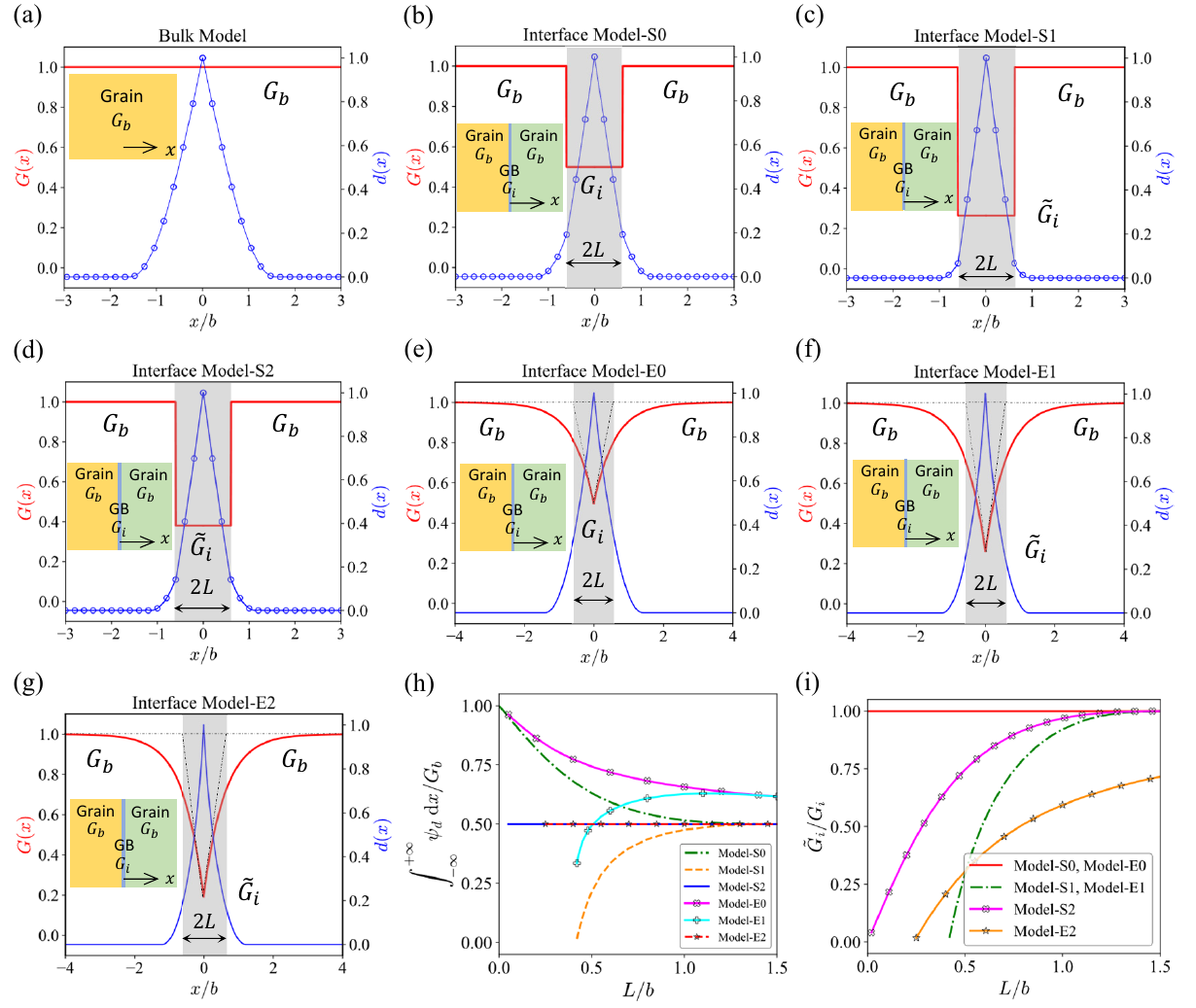}
		\caption{Figure (a) - (g) depict the spatial distributions of fracture energy $G(x)$ and the corresponding phase-field profiles $d(x)$ in the domain in diverse models. 
		Figure (a) illustrates the case with uniform fracture energy in the bulk. 
		Figure (b)-(e) depicts the domain composed of two grains (bulks) and an in-between diffusively represented GB (interface). 
		For aforementioned sub-figures, red line indicates $G(x)$ normalized with bulk fracture energy $G_b$, blue line means $d(x)$ with markers showing its analytical solutions; we set $G_i = 0.5 G_b$ and $L = 0.6b$ when demonstrating diverse interface models.
		Figure (h) shows the integrated phase-field fracture energy w.r.t. $L/b$ in diverse interface models. 
		Figure (i) illustrates obtained effective interface fracture energy $\tilde{G}_i$ normalized with $G_i$, w.r.t. $L/b$ in diverse interface models. 
		$G_i = 0.5 G_b$ is assumed in Figure (h) and (i).
		}
		\label{fig:interface}  
	\end{figure} 

	Consider an 1D infinite long domain in \cref{fig:interface} (a) with $x \in [-\infty, +\infty]$ indicating the spatial coordinate. In the phase-field context, crack evolution is characterized by the phase-field variable $d(x) \in [0, 1]$ describing the material damage level, with $d = 0$ and $d = 1$ denoting sound and fully damaged material, respectively. Follow the paradigm established by \cite{Wu2017JMPS,Miehe2010IJNME}, the bulk crack surface energy density functional $\psi^b_d$ in 1D case in terms of the crack phase-field $d$ and its spatial gradient $d^{\prime}$ is expressed as
	\begin{align}\label{eq:bulk-crack-surface-energy-density}
		\displaystyle{
			\psi^b_d(d,d^{\prime}) = G_b \; \gamma(d , d^{\prime})
		}, \quad
		\text{ with } \quad 
		\gamma(d, d^{\prime}) = \dfrac{1}{\pi} \left( \dfrac{2d - d^2}{b} + b {d^{\prime}}^2 \right),
	\end{align} 
	where $G_b$ represents the critical energy-release rate or is called fracture energy of the bulk material, $b$ is an internal phase-field length-scale parameter regularizing the crack, $2d - d^2$ as the local term expresses the crack geometry function in the cohesive phase-field fracture model \cite{Wu2017JMPS}.

	The fracture energy dissipation of a bulk crack within the phase-field context can be obtained by integrating the density functional \cref{eq:bulk-crack-surface-energy-density} over the domain, i.e., 
	\begin{align}\label{eq:bulk-crack-surface-energy}
		\displaystyle{
			I(d) = \int_{-\infty}^{+\infty} \psi^b_d (d,d^{\prime}) \; \mathrm{d} x,
		}
	\end{align}
	where the phase-field profile $d(x)$ can be determined by minimizing \cref{eq:bulk-crack-surface-energy}, i.e.,
	\begin{align}\label{eq:bulk-crack-surface-energy-minimization}
		\displaystyle{
			d(x) = \text{Arg} \left( \text{inf} \; I(d) \right),
		}
	\end{align}
	leading to following Euler-Lagrange equation associated with the variational problem. Under irreversible evolution constraint \cite{AMM2009JMPS} and proper boundary conditions (assuming the crack to occur in the middle), 
	\begin{align}\label{eq:Euler-Lagrange-bulk}
		\displaystyle{
			\dfrac{\partial \psi^b_d}{\partial d} - \dfrac{\mathrm{d}}{\mathrm{d} x}
			\left( \dfrac{\partial \psi^b_d}{\partial d^{\prime}} \right) = 0,
		}
		\quad \text{subjected to } \;
		\dot{d} \geq 0, \; d(x=0) = 1 \; \text{and} \; d(x= \pm \infty) = 0,
	\end{align}
	phase-field optimal profile reads \cite{Wu2017JMPS}:
	\begin{align}\label{eq:profile-uniform}
		d(x) = 
		\begin{cases}
			1 - \sin \left(\dfrac{|x|}{b}\right), 
			\quad & \text{when }|x| \leq \pi b/2 \\
			0, \quad & \text {else }
		\end{cases}.
	\end{align}
	Accepting the aforementioned spatial distribution of phase-field variable, it becomes evident that integrating the crack surface energy density functional --- implemented through substituting \cref{eq:profile-uniform} into \cref{eq:bulk-crack-surface-energy}, yields the value $G_b$. 
	\begin{align}\label{eq:bulk-crack-equivalence}
		\displaystyle{
			\int_{-\infty}^{+\infty} \psi^b_d(d,d^{\prime})
			\; \mathrm{d} x
			= \int_{-\infty}^{+\infty} \dfrac{G_b}{\pi} \left( \dfrac{2d(x) - d^2(x)}{b} + b {d^{\prime}}^2(x) \right)
			\; \mathrm{d} x
			= G_b.
		}
	\end{align}
	This value accurately quantifies the amount of energy required to create a sharp crack surface of unit area in the bulk material, which effectively remains consistency to the fracture energy defined in the classical cohesive zone model. 
	
	% ***********************************************
	\section{Length-scale insensitivity: Comparison of phase-field interface fracture models}
	\label{sec:interface}

	% with generalizations to prepare for the later multi-dimensional developments. 
	
	In this section, we propose a new phase-field interface model and review several existing models from the literature. Our comparison focuses particularly on verifying the consistency of interface fracture energy with that of the cohesive zone model, as well as examining the sensitivity or insensitivity of the models to length-scale parameters.

	Similar to the setup in previous chapter, 1D infinite long domain is considered with an interface introduced in the middle where the crack is assumed to occur. Sharp interface is diffusively represented in the phase-field context, namely, material properties is interpolated between the bulk phase and the interface; as shown in \cref{fig:interface} (b)-(g) of spatially varying fracture energy $G(x)$ in the domain, which can be formulated in a general form as following,
	\begin{align}\label{eq:G(x)}
		G(x) = A(x) \tilde{G}_i + \left[ 1-A(x) \right] G_b.
	\end{align}	 
	Here $\tilde{G}_i$ denotes the effective interface fracture energy, which can either equal to the sharp interface fracture energy, or be determined from the energy consistency constraint, details can be seen in following diverse interface models; $A(x)$ is the general interpolation function in terms of spatial coordinate.
    %including stairwise and exponential categories widely used in the literatures \cite{Li2020JMPS,Shahed2023TAFM,Gustafsson2022TAFM,Avtar2022JMPS}.

	With introduced spatially-dependent fracture energy $G(x)$, the crack surface energy density functional $\psi_d$ for interface fracture can be constructed,
	\begin{align}\label{eq:interface-crack-surface-energy-density}
		\displaystyle{
			\psi_d(x,d,d^{\prime}) = G(x) \; \gamma(d , d^{\prime})
		}, \quad
		\text{ with } \quad 
		\gamma(d, d^{\prime}) = \dfrac{1}{\pi} \left( \dfrac{2d - d^2}{b} + b {d^{\prime}}^2 \right).
	\end{align} 
	Without loss of generality, as concluded in \cref{table:interface-models} and visualized in \cref{fig:interface} (b)-(g), here we reviewed several existing phase-field interface fracture models (Model-S0, Model-S1, Model-S2, Model-E0 and Model-E1) from the literature and proposed a new model (Model-E2), with particular emphasis on their energy consistency check. Details are as following.

	\begin{sidewaystable}[htbp]
	\centering
	\begin{threeparttable} 
		\caption{Comparison of diverse cohesive phase-field interface models with fracture energy $G(x) = A(x)\tilde{G}_i + (1-A(x))G_b $: check of interface fracture energy and length-scale insensitivity.
		% We set $G_i = 0.5 G_b$ and $L = 0.6b$ when doing energy check in the last column.
		}
		\label{table:interface-models}
		% \begin{tabular}{ m{0pt} m{7cm}<{\centering} m{2cm}<{\centering} m{2.5cm}<{\centering} }
		% \begin{tabular}{|m{2cm}<{\centering} |m{2.5cm}<{\centering} |m{3.5cm}<{\centering} |m{4cm}<{\centering} |m{3cm}<{\centering} | }
		\renewcommand{\arraystretch}{1.5}
		\begin{tabular}{|c|c|c|c|c|c|}
			% -----------------------------------------------
			\hline
				\makecell[c]{
					CPF\\
					interface \\ 
					models }
			& $A(x)$ 
			& \makecell[c]{Effective interface \\fracture energy $\tilde{G}_i$} 
			& $d(x)$			
			& \makecell[c]{
				Interface fracture \\
				energy check \\ 
				$\int_{-\infty}^{+\infty} \psi_d \; {\rm{d}} x \stackrel{?}{=} G_i $ }
			& \makecell[c]{
				Length-scale \\ 
				insensitivity \\
				check ($b, L$)} \\
			% -----------------------------------------------
			% \rule{0pt}{10pt} 
			\hline \xrowht{40pt}
			\makecell[c]{
                Model-S0 \\
                \cite{Gustafsson2022TAFM,Shahed2023TAFM, Chen2024JPS}
                }
			& \makecell[c]{
                $A(x) = H(L-|x|)$\\
                $H$: Heaviside function 
                 \\
            }
			& $\tilde{G}_i = G_i$ 
			& \makecell[l]{
            $ d(x) = 
				\begin{cases}
				1-{\epsilon} \sin \left(\frac{|x|}{b}\right), 
					& \text{when }|x| < L \vspace{-2mm} \\
				1-\sin \left(\frac{|x|}{b}+\theta\right), 
					& \text{when } L \leq|x| \leq\left(\frac{\pi}{2}-{\theta}\right) b \vspace{-2mm} \\
				0, & \text {else } 
				\end{cases} $ \vspace{-1mm} \\
            ${\epsilon}$ and ${\theta}$ determined by $G_b$, ${G}_i$, $b$ and $L$ $^{\dagger}$}
			& $\neq$  
			& sensitive \\
			% -----------------------------------------------
			% \cline{1-1} \cline{3-6} 
			\hline 
			\xrowht{40pt}
			\makecell[c]{
                Model-S1 \\
                \cite{Hansen2019CMAME,Hansen2020EFM,Carlsson2020EFM}
                }
			& $A(x) = H(L-|x|)$
			& \makecell[c]{
                $\tilde{G}_i=\left[G_i-(1-B) G_b\right] / B$ \vspace{1mm} \\
                with $B = \dfrac{1}{\pi} \left[ \dfrac{2L}{b} + \sin \left( \dfrac{2L}{b} \right) \right]$
			}
			& 
			\makecell[l]{
			$ d(x) = \begin{cases}
						1-\tilde{\epsilon} \sin \left(\frac{|x|}{b}\right), 
							& \text{when }|x| < L \vspace{-2mm} \\
						1-\sin \left(\frac{|x|}{b}+\tilde{\theta}\right), 
							& \text{when } L \leq|x| \leq\left(\frac{\pi}{2}-\tilde{\theta}\right) b \vspace{-2mm}\\
						0, & \text {else }
					\end{cases} $ \vspace{-1mm} \\
			$\tilde{\epsilon}$ and $\tilde{\theta}$ determined by $G_b$, $\tilde{G}_i$, $b$ and $L$ $^{\dagger}$}
			& $\neq$ 
			& sensitive \\
			% -----------------------------------------------
			% \cline{1-1} \cline{3-6} 
			\hline 
			\xrowht{68pt}
			\makecell[c]{
                Model-S2 \\
                \cite{Yoshioka2021CMAME,Zhou2022IJSS}
                }
			& $A(x) = H(L-|x|)$
			& \makecell[c]{
				$\tilde{G}_i$ solved from $^{\ast}$ \\
				$\dfrac{L}{b} \dfrac{\tilde{G}_i}{G_b} + \dfrac{\pi}{2}\left(1-\dfrac{G_i}{G_b}\right)$ \vspace{0.5mm} \\
				$-\tan ^{-1}\left[\dfrac{G_b}{\tilde{G}_i} \tan \left( \dfrac{L}{b} \right) \right]=0$
			}
			& 
			\makecell[l]{
			$ d(x) = \begin{cases}
						1-\tilde{\epsilon} \sin \left(\frac{|x|}{b}\right), 
							& \text{when }|x| < L \vspace{-2mm} \\
						1-\sin \left(\frac{|x|}{b}+\tilde{\theta}\right), 
							& \text{when } L \leq|x| \leq\left(\frac{\pi}{2}-\tilde{\theta}\right) b \vspace{-2mm} \\
						0, & \text {else }
					\end{cases} $ \vspace{-1mm} \\
			$\tilde{\epsilon}$ and $\tilde{\theta}$ determined by $G_b$, $\tilde{G}_i$, $b$ and $L$ $^{\dagger}$}
			& $=$ 
			& insensitive \\
			% -----------------------------------------------
			% \rule{0pt}{10pt} &
			\hline \xrowht{40pt}
			\makecell[c]{
                Model-E0 \\
                \cite{Nguyen2016CMAME,Zhang2020CS}
                }
			& $A(x) = \exp \left( -\dfrac{|x|}{L} \right) $
			% \multirow{2}{*}[-1ex]{
			% $ \makecell[l]{
			% 	A(x) = \eta(x) = \exp \left( -\dfrac{|x|}{L} \right) \\
			% 	\text{with } \eta(x) \text{ solved from}\\
			% 	\begin{cases}
			% 		\eta(x)-L^2 \nabla \eta^{\prime \prime}(x)=0 \\
			% 		\eta(0)=1, \; \eta( \pm \infty)=0
		 %        \end{cases}
	  %       } $ }
			& $\tilde{G}_i = G_i$ 
			& \makecell[c]{
				solved from $\dfrac{\partial \psi_d}{\partial d}-\dfrac{\mathrm{d}}{\mathrm{d} x} \left( \dfrac{\partial \psi_d}{\partial d^{\prime}} \right) =0$ with given \\
				$\psi_d = \left[ \exp(-|x|/L) G_i + \left( 1-\exp(-|x|/L) \right) G_b \right] \gamma(d , d^{\prime})$
			} 
			& $\neq$ 
			& sensitive \\
			% -----------------------------------------------
			% \rule{0pt}{10pt} &
			\hline \xrowht{40pt}
			\makecell[c]{
                Model-E1 \\
                \cite{Li2020JMPS,Li2021CMAME}
                }
			& $A(x) = \exp \left( -\dfrac{|x|}{L} \right) $
			% \multirow{2}{*}[-1ex]{
			% $ \makecell[l]{
			% 	A(x) = \eta(x) = \exp \left( -\dfrac{|x|}{L} \right) \\
			% 	\text{with } \eta(x) \text{ solved from}\\
			% 	\begin{cases}
			% 		\eta(x)-L^2 \nabla \eta^{\prime \prime}(x)=0 \\
			% 		\eta(0)=1, \; \eta( \pm \infty)=0
		 %        \end{cases}
	  %       } $ }
			& \makecell[c]{
                $\tilde{G}_i=\left[G_i-(1-B) G_b\right] / B$\\
                with $B = \dfrac{1}{\pi} \left[ \dfrac{2L}{b} + \sin \left( \dfrac{2L}{b} \right) \right]$
			}
			& \makecell[c]{
				solved from $\dfrac{\partial \psi_d}{\partial d}-\dfrac{\mathrm{d}}{\mathrm{d} x} \left( \dfrac{\partial \psi_d}{\partial d^{\prime}} \right) =0$ with given \\
				$\psi_d = \left[ \exp(-|x|/L) \tilde{G}_i + \left( 1-\exp(-|x|/L) \right) G_b \right] \gamma(d , d^{\prime})$
			} 
			& $\neq$ 
			& sensitive \\
			% -----------------------------------------------
			% \cline{1-1}
			% \cline{3-6} 
			\hline
			\xrowht{60pt}
			\makecell[c]{
                Model-E2 \\
                (currently \\ proposed)
                }
			& $A(x) = \exp \left( -\dfrac{|x|}{L} \right) $
			& 
            \multicolumn{2}{c|}{
            \makecell[c]{ 
					$\tilde{G}_i, d(x)$ solved from
					$\begin{cases}
						\int_{-\infty}^{+\infty} \psi_d \; {\rm{d}} x = G_i \vspace{1mm} \\
						\dfrac{\partial \psi_d}{\partial d}-\dfrac{\mathrm{d}}{\mathrm{d} x} \left( \dfrac{\partial \psi_d}{\partial d^{\prime}} \right) =0
					\end{cases}$ \vspace{-1mm} \\
                    with $\psi_d = \left[ \exp(-|x|/L) \tilde{G}_i + \left( 1-\exp(-|x|/L) \right) G_b \right] \gamma(d , d^{\prime})$
			}}
			& $=$
			& insensitive \\
			\hline
			% -----------------------------------------------
			% \bottomrule[1pt]
		\end{tabular}
	\begin{tablenotes} 
		% \item[$\dagger$] \footnotesize{$\tilde{\epsilon} = 1/ \sqrt{\sin^2 \left( \dfrac{L}{b} \right) + \dfrac{\tilde{G}_i}{G_b} \cdot \cos^2 \left( \dfrac{L}{b} \right) }$ and $\tilde{\theta} = \tan ^{-1} \left[ \dfrac{G_b}{\tilde{G}_i} \tan \left( \dfrac{L}{b} \right) \right] - \dfrac{L}{b}$ can be obtained under Weierstrass-Erdmann corner conditions \cite{gelfand2000calculus}, see \cref{appendix-analytical-stairwise} for details.}  
        \item[$^{\dagger}$] \footnotesize{
        $\tilde{\epsilon} = 1/ \sqrt{\sin^2 \left( \dfrac{L}{b} \right) + \dfrac{\tilde{G}_i}{G_b} \cdot \cos^2 \left( \dfrac{L}{b} \right) }$ and $\tilde{\theta} = \tan ^{-1} \left[ \dfrac{G_b}{\tilde{G}_i} \tan \left( \dfrac{L}{b} \right) \right] - \dfrac{L}{b}$ for the corresponding $\tilde{G}_i$ defined in different models. In Model-S0, $\tilde{\epsilon}={\epsilon}, \tilde{\theta}=\theta$ for $\tilde{G}_i=G_i$. 
		}

  	\item[$\ast$] \footnotesize{
		Simplified from $\int_{-\infty}^{+\infty} \psi_d \; {\rm{d}} x = G_i$, with $\psi_d = \left[ H(L-|x|) \tilde{G}_i + \left( 1-H(L-|x|) \right) G_b \right] \gamma(d , d^{\prime})$. 
		}
	\end{tablenotes} 
	\end{threeparttable} 
	% \end{table}
	\end{sidewaystable}

	\subsection{Interface Model-S0: Stair-wise $G(x)$ with $G_i$}\label{sec:model-S0}
	% \begin{itemize}
		% \item \textit{Interface Model-S1 \cite{Gustafsson2022TAFM,Shahed2023TAFM, Chen2024JPS}: Stairwise distribution of $G(x)$ with fracture toughness of a sharp interface $G_i$.}
		As shown in \cref{fig:interface} (b) and formulated in \cref{table:interface-models}, the sharp interface fracture energy $G_i$ is postulated to extend over a finite length $L$ to both sides of the sharp GB \cite{Gustafsson2022TAFM,Shahed2023TAFM, Chen2024JPS}, which can be implemented with the Heaviside step function $H(x)$, i.e.,
		\begin{align}\label{eq:A(x)-stairwise}
			A(x) = H(L-|x|),
			\qquad \text{with} \quad
			H(x) = 
			\begin{cases}
			1, &\quad \text{for } x \geq 0\\
			0, &\quad \text{for } x < 0
			\end{cases},
		\end{align}
		thereby allowing its characterization as a heterogeneous material system featuring two phases: a bulk component with fracture property $G_b$ and a diffusive interface with $G_i$ spanning a width of $2L$. Note the width of interface $2L$ is supposed to be smaller than that where phase-field variable is distributed, in order to eliminate its influence on the fracture properties of bulk materials. 

		Similar to the case in the bulk, with given stair-wise $G(x)$, the profile of phase-field variable $d(x)$ can also be constructed by seeking the optimal one to minimize following integration, 
		\begin{align}\label{eq:interface-crack-surface-energy-minimization}
			\displaystyle{
				d(x) = \text{Arg} \left( \text{inf} \; I(d) \right)
			}, \quad \text{with} \quad
			\displaystyle{
				I(d) = \int_{-\infty}^{+\infty} \psi_d(x,d,d^{\prime})\; \mathrm{d} x.
			}
		\end{align}
		Under above setting, $d(x)$ can be analytically \cite{Yoshioka2021CMAME,Zhou2022IJSS} (see \cref{appendix-analytical-stairwise}) or numerically (with finite-element method) derived from Euler-Lagrange equation of the variational problem additionally subjected to irreversible evolution constraint and proper boundary conditions,
		\begin{align}\label{eq:Euler-Lagrange-interface}
			\displaystyle{
				\dfrac{\partial \psi_d}{\partial d} - \dfrac{\mathrm{d}}{\mathrm{d} x}
				\left( \dfrac{\partial \psi_d}{\partial d^{\prime}} \right) = 0,
			}
			\quad \text{subjected to } \;
			\dot{d} \geq 0, \; d(x=0) = 1 \; \text{and} \; d(x= \pm \infty) = 0,
		\end{align}
		which are formulated as following and visualized in \cref{fig:interface} (b),
		\begin{align}\label{eq:profile-Model-S0}
			d(x) = 
			\begin{cases}
				1-\epsilon \sin \left(\dfrac{|x|}{b}\right), 
					\quad & \text{when }|x| \leq L \\
				1-\sin \left(\dfrac{|x|}{b}+\theta\right), 
					\quad & \text{when } L < |x| \leq\left(\dfrac{\pi}{2}-\theta\right) b \\
				0, \quad & \text {else }
			\end{cases}, 
			\quad \text{with} \quad
			\begin{cases}
				\epsilon = 1/ \sqrt{\sin^2 \left( \dfrac{L}{b} \right) + \left( \dfrac{G_i}{G_b} \right)^2 \cdot \cos^2 \left( \dfrac{L}{b} \right) } \vspace{2mm} \\
				\theta = \tan ^{-1} \left[ \dfrac{G_b}{G_i} \tan \left( \dfrac{L}{b} \right) \right] - \dfrac{L}{b}
			\end{cases}.
		\end{align}

		By substituting above profile result $d(x)$ into crack surface energy density functional \cref{eq:interface-crack-surface-energy-density}, we obtain following integrated phase-field interface fracture energy, 
		\begin{align}\label{eq:integral-Model-S0}
			\displaystyle{
				\int_{-\infty}^{+\infty} 
				% G(x) \gamma(d , d^{\prime})
				\psi_d (x,d,d^{\prime})
				\; \mathrm{d} x
				= \dfrac{2}{\pi} \left[
				\left( \dfrac{\pi}{2} - \tan ^{-1} \left[ \dfrac{G_b}{G_i} \tan \left( \dfrac{L}{b} \right) \right] + \dfrac{L}{b} \right) G_b + 
				\left( G_i - G_b \right) \dfrac{L}{b}
				\right] 
				\neq G_i.
			}
		\end{align}
		As visualized in \cref{fig:interface} (h) and also evidenced above and in \cref{table:interface-models}, the obtained integration result does not align with the sharp interface fracture energy. Consequently, Model-S0 has failed to accurately assess the fracture property of a sharp interface.
		Besides, according to \cref{eq:integral-Model-S0} plotted in \cref{fig:interface} (h), the integrated fracture energy within the phase-field framework is dependent on length-scale parameters $b$ and $L$, e.g., fracture property of the interface is significantly overestimated when $L/b$ is smaller, so that numerical outcomes including fracture patterns and structure reaction forces are expected to be sensitive to the choice of length-scale parameters. This sensitivity is highlighted in the results of examples presented in \cref{results}. 

	\subsection{Interface Model-S1: Stair-wise $G(x)$ with $\tilde{G}_i$} \label{sec:model-S1}

		In order to constrain the integrated phase-field fracture energy consistent with the sharp interface property, \cite{Hansen2019CMAME,Hansen2020EFM} proposed to use the effective interface fracture energy, as can be seen in \cref{fig:interface} (c) with $\tilde{G}_i$ extending a finite distance $L$ from the sharp GB to its both sides.

		A proper effective value of $\tilde{G}_i$ can be determined to ensure the interface fracture energy consistent with the sharp one, i.e., 
		\begin{align}\label{eq:equation-Model-S1}
			\displaystyle{
				\int_{-\infty}^{+\infty} 
				% G(x) \gamma(d , d^{\prime})
				\psi_d (x,d,d^{\prime})
				\; \mathrm{d} x
				=  G_i.
			}
		\end{align}	
		To solve the above equation, the phase-field profile $d(x)$ in \cref{eq:profile-uniform} for homogeneous material with spatially uniform fracture energy is assumed in \cite{Hansen2019CMAME,Hansen2020EFM}. With direct derivation, following expression of $\tilde{G}_i$ in terms of length-scale parameters can be obtained,
		\begin{align}\label{eq:Gi-tilde-Molde-S1}
			\tilde{G}_i=\left[G_i-(1-B) G_b\right] / B, \quad \text{with} \quad
			B = 2 \int_0^L \gamma \left( d, d^{\prime} \right) {\mathrm{d}} x
			= \dfrac{1}{\pi} \left[ \dfrac{2L}{b} + \sin \left( \dfrac{2L}{b} \right) \right].
		\end{align}
		which is visualized in \cref{fig:interface} (i) w.r.t. $L/b$.

		However, for a heterogeneous material system including both bulk and diffusive interface, it is not appropriate to assume the phase-field profile in \cref{eq:equation-Model-S1} identical to that under spatially uniform fracture property; instead, the phase-field profile is supposed to be obtained by solving the Euler-Lagrange equation expressed in \cref{eq:Euler-Lagrange-interface}, so that $d(x)$ yields,
		\begin{align}\label{eq:profile-Model-S1}
			d(x) = 
			\begin{cases}
				1-\tilde{\epsilon} \sin \left(\dfrac{|x|}{b}\right), 
				\quad & \text{when }|x| \leq L \\
				1-\sin \left(\dfrac{|x|}{b}+\tilde{\theta}\right), 
				\quad & \text{when } L < |x| \leq\left(\dfrac{\pi}{2}-\tilde{\theta}\right) b \\
				0, \quad & \text {else }
			\end{cases}, 
			\quad \text{with} \quad
			\begin{cases}
				\tilde{\epsilon} = 1/ \sqrt{\sin^2 \left( \dfrac{L}{b} \right) + \left( \dfrac{\tilde{G}_i}{G_b} \right)^2 \cdot \cos^2 \left( \dfrac{L}{b} \right) } \vspace{2mm} \\
				\tilde{\theta} = \tan ^{-1} \left[ \dfrac{G_b}{\tilde{G}_i} \tan \left( \dfrac{L}{b} \right) \right] - \dfrac{L}{b}
			\end{cases}.
		\end{align}
		where the effective interface fracture energy $\tilde{G}_i$ is expressed in \cref{eq:Gi-tilde-Molde-S1}. To continue, we can substitute the optimal phase-field profile result $d(x)$ into the integration of crack surface energy density functional \cref{eq:interface-crack-surface-energy-density},
		\begin{align}\label{eq:integral-Model-S1}
			\displaystyle{
				\int_{-\infty}^{+\infty} 
				% G(x) \gamma(d , d^{\prime})
				\psi_d (x,d,d^{\prime})
				\; \mathrm{d} x
				= \dfrac{2}{\pi} \left[
				\left( \dfrac{\pi}{2} - \tan ^{-1} \left[ \dfrac{B G_b}{G_i - (1-B)G_b} \tan \left( \dfrac{L}{b} \right) \right] + \dfrac{L}{b} \right) G_b + 
				\dfrac{G_i - G_b}{B} \dfrac{L}{b}
				\right] 
				\neq G_i,
			}
		\end{align}
		with $B$ defined in \cref{eq:Gi-tilde-Molde-S1}.

		As can be concluded from \cref{fig:interface} (h) and in \cref{table:interface-models}, the above integrated phase-field interface fracture energy in Model-S1 can not achieve the consistency with the fracture property of a sharp interface, which tends to underestimate the structure resistance of failure. Besides, due to imprecise assess of the interface fracture property, numerical results predicted by Model-S1 are also expected to have the issue of length-scale sensitivity.

	% \item \textit{Interface Model-S2 \cite{}: Stairwise distribution of $G(x)$ with effective interfacial fracture toughness $\tilde{G}_i$, subjected to crack surface energy dissipation equivalence at the interface.}
	\subsection{Interface Model-S2: Stair-wise $G(x)$ with $\tilde{G}_i$ and interface fracture energy consistency} \label{sec:model-S2}

		Instead of using the phase-field profile in \cref{eq:profile-uniform} under spatially uniform fracture energy, \cite{Yoshioka2021CMAME,Zhou2022IJSS} analytically solve the phase-field profile $d(x)$ from the Euler-Lagrange equation and express the result w.r.t. the effective interface fracture energy to be determined,
		\begin{align}\label{eq:profile-Model-S2}
			d(x) = 
			\begin{cases}
				1-\tilde{\epsilon} \sin \left(\dfrac{|x|}{b}\right), 
				\quad & \text{when }|x| \leq L \\
				1-\sin \left(\dfrac{|x|}{b}+\tilde{\theta}\right), 
				\quad & \text{when } L < |x| \leq\left(\dfrac{\pi}{2}-\tilde{\theta}\right) b \\
				0, \quad & \text {else }
			\end{cases}, 
			\quad \text{with} \quad
			\begin{cases}
				\tilde{\epsilon} = 1/ \sqrt{\sin^2 \left( \dfrac{L}{b} \right) + \left( \dfrac{\tilde{G}_i}{G_b} \right)^2 \cdot \cos^2 \left( \dfrac{L}{b} \right) } \vspace{2mm} \\
				\tilde{\theta} = \tan ^{-1} \left[ \dfrac{G_b}{\tilde{G}_i} \tan \left( \dfrac{L}{b} \right) \right] - \dfrac{L}{b}
			\end{cases}.
		\end{align}
		With explicitly expressed phase-field profile $d(x)$ under stair-wise $G(x)$, constraining equation that ensures the integrated phase-field fracture energy equivalent to that of sharp interface can be further simplified so that $\tilde{G}_i$ can be solved, namely,
		\begin{align}\label{eq:equation-model-S2}
			\displaystyle{
				\int_{- \infty}^{+ \infty} \psi_d (x,d,d^{\prime})
					\; \mathrm{d} x
				= G_i
			}
			\qquad
			\Longrightarrow
			\qquad
			\dfrac{L}{b} \dfrac{\tilde{G}_i}{G_b} - 
				\tan^{-1} \left[\frac{G_b \tan \left( \frac{L}{b} \right)}{\tilde{G}_i} \right] + 
					\dfrac{\pi}{2} \left( 1 - \dfrac{G_i}{G_b} \right) = 0.
		\end{align} 

		A visualization of obtained $\tilde{G}_i$ from above equation in terms of given length-scale parameters $b$ and $L$ can be seen in \cref{fig:interface} (i) under the condition $G_i = 0.5G_b$. \cref{fig:interface} (d) depicts the phase-field profile $d(x)$ in \cref{eq:profile-Model-S2} under stair-wise $G(x)$ with solved $\tilde{G}_i$. \cref{fig:interface} (h) shows the agreement between the integrated phase-field fracture energy and the sharp value in the cohesive zone model. It can be concluded that, with elaborately determined effective interface fracture energy $\tilde{G}_i$, energetic equivalence between the sharp and diffused representation of interface fracture can be checked and ensured accurately, despite of varying choices of length-scale parameters, which accordingly promises numerical results independent on the choices of length-scale parameters.
		
	\subsection{Interface Model-E0: Exponential $G(x)$ with $G_i$}\label{sec:model-E0}

		In this model, the interpolation function $A(x)$ is constructed on the basis of an auxiliary (phase-field) indicator $\eta(x)$ characterizing the interface configuration, e.g., simplest case $A(x) = \eta(x)$ is considered, though more complicated forms like quadratic one \cite{Zhang2019EFM} could also be adopted. The following Allen-Cahn type governing equation and boundary conditions \cite{Nguyen2016CMAME} are needed to solve the spatial distribution of the indicator $\eta(x)$,
		\begin{align}
			\begin{cases}
				\eta(x)-L^2 \eta^{\prime \prime}(x)=0, \quad
				& \text{in } \varOmega \\
				\nabla \eta \cdot \bfn = 0, \quad & \text{on } \partial \varOmega 
				% \eta(0)=1, \; \eta( \pm \infty)=0
				% & \quad \tilde{G}_i=G_i, \; \eta(x)=\exp \left( -\frac{|x|}{L} \right)
			\end{cases},
			\qquad \text{subjected to } \; \eta(x=0)=1, \; \eta(x= \pm \infty)=0,
		\end{align}
		where $\varOmega$ denotes the 1D domain whose external boundary $\partial \varOmega$ has outward unit normal vector $\bfn$; $L$ is the length-scale parameter regularizing the interface and it controls the varying slope of interface indicator; $x = 0$ represents the location of interface and $x = \pm \infty$ are two far ends of the domain. Its analytical solution in 1D domain is written as:
		\begin{align}
			\eta(x) = \exp(-|x|/L).
		\end{align}
		With the interface indicator $\eta(x)$ at hands, the fracture energy in the domain can be formulated as \cite{Nguyen2016CMAME,Zhang2020CS}
		\begin{align}
			A(x) = \eta(x) \qquad \Longrightarrow \qquad G(x) = \exp \left(-\dfrac{|x|}{L} \right) G_i + \left( 1 - \exp \left(-\dfrac{|x|}{L} \right) \right)G_b,
		\end{align}
		as shown in \cref{fig:interface} (d), the center exhibits the sharp interface fracture energy $G_i$, smoothly transitioning to the bulk property $G_b$ via an exponential function as depicted above.

		With known spatial distribution of fracture energy $G(x)$, phase-field profile $d(x)$ can be solved from Euler-Lagrange equation additionally subjected to irreversible evolution constraint and proper boundary conditions, see \cref{eq:Euler-Lagrange-interface} for the equation and \cref{fig:interface} (e) for the visualization of its numerical solution. Given the intricate exponential distribution of $G(x)$, it's important to note that there exists no analytical expression for $d(x)$, unlike previous models characterized by a stair-wise $G(x)$. 

		Subsequently, integrated phase-field fracture energy can be determined and checked with that of a sharp interface,
		\begin{align}
			\displaystyle{
				\int_{-\infty}^{+\infty} 
				% G(x) \gamma(d , d^{\prime})
				\psi_d (x,d,d^{\prime})
				\; \mathrm{d} x
				= \int_{-\infty}^{+\infty} \dfrac{G(x)}{\pi} \left( \dfrac{2d(x) - d^2(x)}{b} + b {d^{\prime}}^2(x) \right)
		\; \mathrm{d} x
				\neq G_i
			}.
		\end{align}
		As shown in \cref{fig:interface} (h) with numerical integration result, Model-E0 falls short in accurately evaluating the crack surface energy for interface fracture, which overestimates the fracture property of the interface like Model-S0, and might lead to imprecise predictions of crack propagation paths and corresponding mechanical responses. Moreover, the parameters $b$ and $L$ significantly influence the determination of the phase-field profile $d(x)$ and the corresponding integration result of crack surface energy, resulting in notable dependency and sensitivity of simulation results on length-scale parameters, which can be seen in numerical examples in \cref{results}. 

		% However, aforementioned shortcoming is always ignored in the literatures, including both purely-mechanical \cite{Li2020JMPS,Gustafsson2022TAFM} and chemo-mechanical \cite{Avtar2022JMPS,Emilio2023JMCA} scenarios.

	\subsection{Interface Model-E1: Exponential $G(x)$ with $\tilde{G}_i$} \label{sec:model-E1}

		On the basis of the setting in Model-E0, \cite{Li2020JMPS,Li2021CMAME} proposed to employ an effective interface fracture energy in the exponential $G(x)$, specifically, the effective value of interface fracture energy $\tilde{G}_i$ is determined according to Model-S1 \cite{Hansen2019CMAME}, i.e., expressed in \cref{eq:Gi-tilde-Molde-S1}, which smoothly transitions to the bulk fracture energy $G_b$ with an exponential function, as displayed in \cref{fig:interface} (f).
		\begin{align}\label{eq:G(x)-model-E1}
		\displaystyle{
			G(x) = \exp \left(-\dfrac{|x|}{L} \right) \tilde{G}_i + \left( 1 - \exp \left(-\dfrac{|x|}{L} \right) \right)G_b
		}.
		\end{align}

		In order to implement the interface fracture energy check under above setting, firstly the phase-field profile $d(x)$ is numerically determined (with finite-element method) from the Euler-Lagrange equation (\cref{eq:Euler-Lagrange-interface}) under given $G(x)$, with results visualized in \cref{fig:interface} (f). Subsequently, by substituting above profile result into \cref{eq:interface-crack-surface-energy-density}, integrated phase-field fracture energy can be calculated w.r.t varying length-scale parameters $(L,b)$ and illustrated in \cref{fig:interface} (h),
		\begin{align}
			\displaystyle{
				\int_{-\infty}^{+\infty} 
				% G(x) \gamma(d , d^{\prime})
				\psi_d (x,d,d^{\prime})
				\; \mathrm{d} x
				= \int_{-\infty}^{+\infty} \dfrac{G(x)}{\pi} \left( \dfrac{2d(x) - d^2(x)}{b} + b {d^{\prime}}^2(x) \right)
		\; \mathrm{d} x
				\neq G_i
			}.
		\end{align}
		As can be concluded above and in \cref{table:interface-models}, Model-E1 also fails to assess the interface fracture energy, which differs from the sharp value and exhibits dependency under varying choice of length-scale parameters. Such discrepancy and length-scale sensitivity can also be observed in the numerical examples in \cref{results}.

	\subsection{Interface Model-E2: Exponential $G(x)$ with $\tilde{G}_i$ and interface fracture energy consistency} \label{sec:model-E2}

		In order to address the aforementioned shortcomings in Model-E0 and Model-E1, inspired by Model-S2 and for the first time, we employ the effective interface fracture energy $\tilde{G}_i$ in the exponential $G(x)$, which is solved on the basis of the Euler-Lagrange equation of the phase-field theory and the interface energy consistency w.r.t. that of the cohesive zone model.
		
		Similar to Model-E1, the fracture energy distribution is given as following,
		\begin{align}\label{eq:G(x)-model-E2}
			\displaystyle{
				G(x) = \exp \left(-\dfrac{|x|}{L} \right) \tilde{G}_i + \left( 1 - \exp \left(-\dfrac{|x|}{L} \right) \right)G_b
			},
		\end{align}
		where the effective interface fracture energy $\tilde{G}_i$ can be determined by constraining the fracture energy consistent with that defined in the cohesive zone model. Meanwhile, the phase-field profile $d(x)$ can be solved according to the Euler-Lagrange equation under above $G(x)$. Accordingly, the unknown scalar $\tilde{G}_i$ and the unknown variable $d(x)$ can be simultaneously determined by solving following coupled equations,
		\begin{align}\label{eq:equation-model-E2}
			\begin{cases}
				\displaystyle{
				\int_{-\infty}^{+\infty} 
				% G(x) \; \gamma\left(d, d^{\prime}\right) 
				\psi_d(x,d,d^{\prime}) \; 
				{\rm{d}} x = G_i } \vspace{2mm} \\
				\displaystyle{
					\dfrac{\partial \psi_d}{\partial d} - \dfrac{\mathrm{d}}{\mathrm{d} x}
					\left( \dfrac{\partial \psi_d}{\partial d^{\prime}} \right) = 0,
				}
				\quad \text{subjected to } \;
				\dot{d} \geq 0, \; d(x=0) = 1 \; \text{and} \; d(x= \pm \infty) = 0
			\end{cases}.
		\end{align}
		As plotted in \cref{fig:interface} (g), the spatial distribution of fracture energy $G(x)$ with $\tilde{G}_i$ and phase-field variable $d(x)$ are illustrated. Besides, $\tilde{G}_i$ obtained from above equation system in terms of varying length-scale parameters $b$ and $L$ under the condition $G_i = 0.5G_b$ is illustrated in \cref{fig:interface} (i).

		% Similarly, optimal profile $d(x)$ can be solved from \cref{eq:Euler-Lagrange-non-uniform} using finite-element method with given exponential $G(x)$, so that check of energy dissipation equivalence  in \cref{eq:integral-Model-S1} can be made. 
		\cref{fig:interface} (h) plotted the integrated phase-field fracture energy under varying $L/b$, which is constrained with energy equivalence to that of a sharp interface. It can be concluded that, Model-E2, which employs a diffusive exponential representation of the interface additionally equipped with an elaborately determined effective interface fracture energy, establishes equivalence to a sharp interface crack in terms of fracture energy dissipation. Consequently, the proposed model anticipates the advantageous insensitivity to length-scale parameters ($b$ and $L$). 
	% \end{itemize}

	% \remark \label{remark:interface-fracture-energy-check}
	% 	% 
	% 	\begin{table}[htbp]
	% 	\centering
	% 	\caption{Comparison of ``Interface Fracture Energy Check" among various phase-field interfaces models, assuming $G_i = 0.5 G_b$ and $L = 0.6b$.}
	% 	\label{table:interface-energy-check}
	% 	\begin{tabular}{|c|c|c|c|c|}
	% 		% 
	% 		\hline \xrowht{15pt}
	% 			Interface Models
	% 			& Model-S1
	% 			& Model-S2
	% 			& Model-E1
	% 			& Model-E2 \\
	% 		% 
	% 		\hline \xrowht{20pt}
	% 			$\int_{-\infty}^{+\infty} 
	% 			% G(x) \; \gamma\left(d, d^{\prime}\right) 
	% 			\psi_d(x,d,d^{\prime}) 
	% 			{\rm{d}} x \stackrel{?}{=} G_i $ 
	% 			& $ 0.59 G_b \neq G_i$					
	% 			& $ 0.50 G_b = G_i$					
	% 			& $ 0.72 G_b \neq G_i$					
	% 			& $ 0.50 G_b = G_i$ \\
	% 		% 
	% 		\hline
	% 	\end{tabular}
	% 	\end{table}
	% 	% 
	% 	In order to illustrate ``interface fracture energy check" of various phase-field interfaces models, here we set $G_i = 0.5 G_b$ and let $L = 0.6b$ to calculate the crack surface energy numerically, and make comparisons between that of a sharp interface. With given conditions, as listed in \cref{table:interface-energy-check}, it can be concluded that Model-S2 and Model-E2 can accurately realize the replication of fracture energy dissipation of a sharp interface, while the other two models failed. 
	% $\Box$
	\remark \label{remark:numerical-solving}
		In order to solve the equation system in \cref{eq:equation-model-E2}, which includes a Transcendental Equation (TE) for the effective interface fracture energy $\tilde{G}_i$ and a Partial Differential Equation (PDE) for the phase-field variable $d(x)$, we adopt a staggered algorithm to sequentially and iteratively solve two separate equations, e.g., initial value $\tilde{G}_i=G_i$ is set to start the iteration, then profile $d(x)$ can be obtained by solving the PDE using finite-element method under given $G(x)$, next, the bisection method properly bisecting the original interval $\left[ 0,\tilde{G}_i\right]$ is applied to solve the TE. Aforementioned procedures consisting of staggered iterations and bisecting the interval are repeated, until the length of the updated interval for $\tilde{G}_i$ in the bisection method meets the defined threshold. 
	$\Box$

	\remark \label{remark:more-diffusive-representation}
		It is noteworthy to highlight that the diffusive representation of an interface, incorporating an elaborately determined effective interface fracture energy, is capable of ensuring the fracture energy consistent with that of a sharp interface crack. Moreover, this applicability is not confined to the specific types of interpolation as previously discussed, e.g., stair-wise and exponential $G(x)$, as it extends to encompass a more general representation of the interface, such as the parabolic one in \cite{Lorenzis2023EJM}, which can be constructed by properly defining $A(x)$ in \cref{eq:G(x)}.
	$\Box$

	\remark \label{remark:unified-density-bulk-interface}
		As for the crack surface energy density functional $\psi_d(x,d,d^{\prime})$ in \cref{eq:interface-crack-surface-energy-density} for interface fracture, when the fracture energy of bulk material is assigned over the whole domain, i.e., $G(x) = G_b$, the density functional $\psi^{b}_d(d,d^{\prime})$ in \cref{eq:bulk-crack-surface-energy-density} for the bulk fracture can be recovered. In the following chapter for multi-dimensional and multi-physical development, $\psi_d(x,d,d^{\prime})$ in the unified form is employed, which can model fractures both within the bulk and at the interface.
	$\Box$

	\section{Chemo-mechanical cohesive phase-field model for inter- and trans-granular fractures}
	\label{sec:phase-field}
	On the basis of the proposed phase-field interface model, in this section, we extend its applicability to chemo-mechanical cohesive fracture in general multi-dimensional case in a thermodynamically consistent manner, which can apply to both inter-granular fracture at the grain boundary and trans-granular fracture within the grain.
	% Interested readers are suggested to refer to \cite{Wu2017JMPS,Chen2024JPS} for detailed derivations.

	\begin{figure}[htbp]
		\centering
		\includegraphics[width=\textwidth]{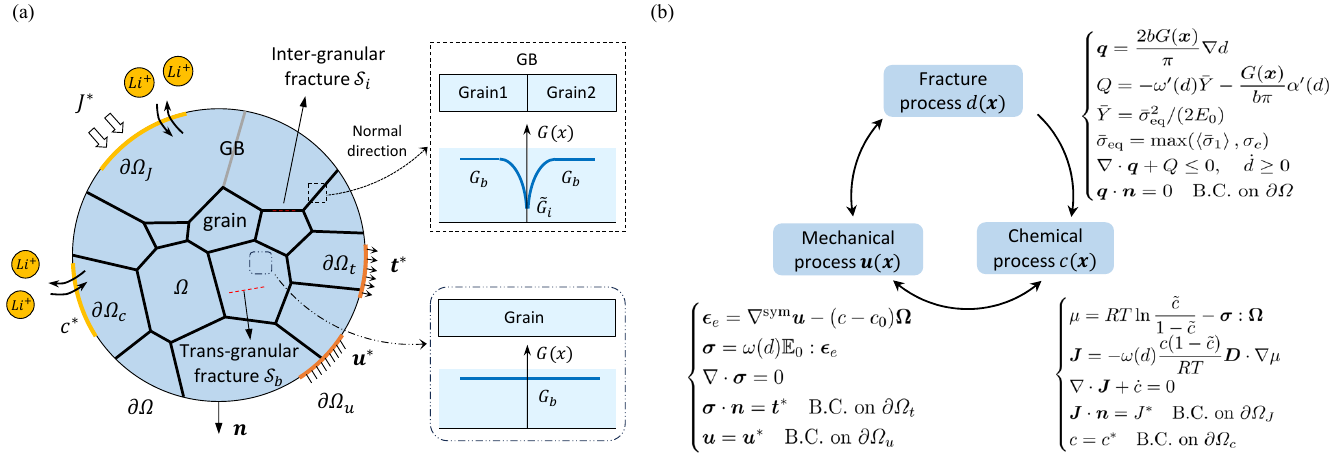}
		\caption{Illustrations of cohesive phase-field model for chemo-mechanical fractures in a NMC polycrystalline particle. Figure (a) presents a typical microstructure of NMC particle with multiple grains and GBs. The boundary conditions for both chemical and mechanical sub-problems are illustrated, possible crack patterns including inter- and trans- granular fracture are also provided. The distribution of the fracture energy $G(x)$ in the diffusively represented GB (utilizing Model-E2) and the bulk material are highlighted in the zoomed call-out. Although this illustration is shown in 2D, models in this chapter are applicable to 3D simulations. Figure (b) shows the chemical-mechanical-fracture coupled formulations and the couplings of different physics.}
		\label{fig:configuration}
	\end{figure}
	
	As shown in \cref{fig:configuration} (a), let $\varOmega \subset \mathbb{R}^{n_{\dim}}$ ($n_{\dim} = 1, 2, 3$) be the reference configuration of heterogeneous polycrystalline solid with multiple grains and GBs, its external boundary is denoted by $\partial \varOmega \subset \mathbb{R}^{n_{\dim} - 1}$ with the outward normal vector $\boldsymbol{n}$. A multi-dimensional extension of the crack phase-field variable reads $d(\bfx) : \varOmega \in [0, 1]$, with corresponding crack surface energy density functional in multi-dimensional case defined as following,
	\begin{align}\label{eq:crack-surface-energy-density-3D}
		\displaystyle{
			% \int_{\varOmega} 
			\psi_d (\bfx, d, \nabla d) = 
			 % \; \mathrm{d} V = 
			% 
			% \int_{\varOmega} 
			G(\bfx) \; \gamma(d , \nabla d)
			% \; \mathrm{d} V
		}, \quad
		\text{ with } \quad 
		\gamma(d, \nabla d) = \dfrac{1}{\pi} \left( \dfrac{2d - d^2}{b} + b |\nabla d|^2 \right),
	\end{align}
	where $\gamma(d, \nabla d)$ in terms of phase-field variable and its spatial gradient expresses the regularized crack surface density functional, $G(\bfx)$ is the fracture energy of material point distinguishing the bulk phase and the diffusive interface, with $\bfx$ labeling the spatial coordinates, see \cref{fig:configuration} (a) for the illustration. In 2D/3D cases, the distribution of fracture energy $G(x)$ along the normal direction of GBs with elaborately determined $\tilde{G}_i$ and the corresponding phase-field profile $d(x)$ adhere to the configurations established in the preceding 1D case discussed in the previous chapter, so that fracture energy dissipated for inter-granular cracks at the sharp GBs can be precisely recaptured within the phase-field framework. In the whole domain, we have, 
    \begin{align}\label{eq:3D-energy-consistence}
    \int_{\varOmega} \psi_d(\bfx, d, \nabla d) \; \mathrm{d} V = \int_{\mathcal{S}_i} G_i \; \mathrm{d} A  
    + \int_{\mathcal{S}_b} G_b \; \mathrm{d} A,
    \end{align}
    which covers the energy dissipation for concurrent bulk and interface fractures. Refer to \cref{fig:configuration} (a) for a zoomed call-out, which provides an illustrative depiction of $G(x)$ along the normal direction of a GB, accompanied by potential inter-granular (interface) crack $\mathcal{S}_i$, as well as the bulk material with uniform fracture energy $G_b$ demonstrating potential trans-granular (bulk) crack $\mathcal{S}_b$. 
    
    Under the above settings, we will present the thermodynamic derivations for a chemo-mechanically coupled cohesive phase-field model for concurrent bulk and interface fracture in following subsections.

	% \subsection{Formulations under purely-mechanical circumstance}
	\subsection{Kinematics and chemo-mechanical governing equations}

	As also can be seen in \cref{fig:configuration} (a), the chemo-mechanical behaviors of NMC particle are characterized by the displacement (vector) field $\boldsymbol{u} (\boldsymbol{x}, t)$ and the concentration (scalar) field $c (\boldsymbol{x}, t)$, in addition to phase-field crack $d(\bfx)$, with $\boldsymbol{x}$ and $t$ labeling the material point and time, respectively. 

	Under the small-deformation setting, the total strain tensor $\bfepsilon (\boldsymbol{x})$ is obtained via the symmetric part of the displacement gradient tensor, which can be further decomposed into two parts, i.e., elastic $\bfepsilon_{e}$ and chemical $\bfepsilon_{c}$ strain tensors.
	\begin{subequations}\label{eq:kinematics}
		\begin{align}
			\label{eq:kinematics-total}
			&
			\bfepsilon = \nabla ^{\text{sym}} \boldsymbol{u} = \dfrac{1}{2} (\nabla \boldsymbol{u} + \nabla^{\rm{T}} \boldsymbol{u}) = 
				\bfepsilon_{e} + \bfepsilon_{c},
			\\
			\label{eq:kinematics-chemical}
			&
			\bfepsilon_{c} = (c - c_0) \boldsymbol{\Omega}, 
				\qquad \boldsymbol{\Omega} = \Omega_{ij} \boldsymbol{e}_{i} \otimes \boldsymbol{e}_{j}.
		\end{align}
	\end{subequations}
	The elastic part directly contributes to the mechanical stress, while the chemical one measures the material volume change induced by varying concentration within the host material during (dis)charging processes. Here $c_0$ denotes the initial concentration in the active material under stress-free state; $\boldsymbol{\Omega}$ is the chemical deformation (swelling/shrinkage) tensor determining lithium-concentration-dependent volume change, its components are represented by the partial molar volume $\Omega_{ij}$. 
	% $\boldsymbol{\Omega}$ is described by a transversely isotropic model \cite{} to distinguish different diffusion abilities between in-plane ($a-b$) and out-of-plane ($c$) directions, as illustrated in \cref{fig:example3} (a) of the micro structure and crystal orientations.

	In the mechanically quasi-static case (body force is ignored) under consideration, the mechanical stress tensor $\bfsigma$ and the chemical flux $\bfJ$ in the current configuration are governed by the following equations:
	\begin{subequations}\label{eq:governing-equations}
		\begin{align}
			\label{eq:governing-equations-mechanical}
			&\begin{cases}
				\nabla \cdot \bfsigma = \boldsymbol{0} 
				\qquad \quad &
				\text{in} \; \varOmega \\
				\bfsigma \cdot \boldsymbol{n} = \boldsymbol{t}^{\ast}
				\qquad \quad &
				\text{on} \; \partial \varOmega_{t} \\
				\bfu = \bfu^{\ast}
				\qquad \quad &
				\text{on} \; \partial \varOmega_{u} \\
			\end{cases}, \\
			\label{eq:governing-equations-chemical}
			&\begin{cases}
				\nabla \cdot \boldsymbol{J} + \dot{c} = 0
				\qquad \quad &
				\; \text{in} \; \varOmega \\
				\boldsymbol{J} \cdot \boldsymbol{n} = J^{\ast} 
				\qquad \quad &
				\; \text{on} \; \partial \varOmega_{J} \\
				c = c^{\ast}
				\qquad \quad &
				\; \text{on} \; \partial \varOmega_{c} \\
			\end{cases}.
		\end{align}
	\end{subequations}
	Here prescribed traction force $\boldsymbol{t}^{\ast}$ (on $\partial \varOmega_{t}$) and displacement $\boldsymbol{u}^{\ast}$ (on $\partial \varOmega_{u}$) are boundary conditions on the external surface $\partial \varOmega$ for mechanical sub-problem; chemical flux $J^{\ast}$ and concentration $c^{\ast}$ applied on surfaces $\partial \varOmega_{J}$ and $\partial \varOmega_{c}$, respectively, serve as chemical boundary conditions, in order to mimic the electrochemical responses and lithium (de)intercalation flux on the interface between electrolyte and electrode particles \cite{Zhao2016CMAME,Shahed2023TAFM}.
	\subsection{Thermodynamic and dissipation inequality}

	Under chemo-mechanical-fracture coupled circumstance, the constitutive laws of all processes and the governing equation of crack phase-field are derived in a thermodynamically consistent manner.

	Consider an isothermal and adiabatic system, the second law of thermodynamics dictates that, in its global formulation, the interplay between external power and internal potential energy entails an energy dissipation rate that demands non-negative, i.e., 
	\begin{align}\label{eq:energy-dissipation}
		\dot{\mathscr{D}} = 
		\int_{\partial \varOmega} \boldsymbol{t}^{\ast} \cdot \dot{\bfu} \; \mathrm{d} A -
		% \int_{\varOmega} \boldsymbol{b}^{\ast} \cdot \dot{\bfu} \mathrm{d} V +
		\int_{\partial \varOmega} J^{\ast} \mu \; \mathrm{d} A -
		\int_{\varOmega} \dot{\psi} \; \mathrm{d} V 
		% 
		% = \int_{\varOmega} \left(
		% \bfsigma : \dot{\bfepsilon} % mechanical external power
		% + \mu \dot{c} - \bfJ \cdot \nabla \mu 
		% -\dot{\psi} % potential
		% \right) \mathrm{d} V 
		% 
		\geq 0,
	\end{align}
	where time derivation is denoted by $\dot{()}$, $\mu$ is the chemical potential, and $\psi$ represents internal potential energy density taking into account the contributions from all processes including chemical, mechanical and fracture sub-problems,
	\begin{subequations}\label{eq:potential-energy}
		\begin{align}
			\label{eq:potential-energy-total}
			&
			\psi = \psi_c(c, \nabla c)
			+ \psi_m(\bfepsilon (\bfu), c, d)
			+ \psi_d(\bfx, d, \nabla d),
			\\
			\label{eq:potential-energy-chemical}
			&
			\psi_c = R T c_{\max } [\tilde{c} \ln \tilde{c}+(1-\tilde{c}) \ln (1-\tilde{c})],
			\\
			\label{eq:potential-energy-mechanical}
			&
			\psi_m = \dfrac{1}{2} \bfepsilon_e: \omega (d) \mathbb{E}_0: \bfepsilon_e
			= \dfrac{1}{2} \left( \bfepsilon - \bfepsilon_c \right) 
			: \omega (d) \mathbb{E}_0 : \left( \bfepsilon - \bfepsilon_c \right),
			% \\
			% % 
			% \label{eq:potential-energy-damage}
			% &
			% \psi_d = G (\boldsymbol{x}) \; \gamma(d, \nabla d) = \dfrac{G (\boldsymbol{x})}{\pi} \left( \dfrac{2d - d^2}{b} + b |\nabla d|^2 \right)
			% \\
		\end{align}
	\end{subequations}
	together with crack surface energy density functional $\psi_d(\bfx, d, \nabla d)$ defined in \cref{eq:crack-surface-energy-density-3D}. In above formulations, $R$ is the gas constant, $T$ is the reference temperature, $c_{\max}$ denotes the maximum Li concentration in active material for normalizing Li concentration $\tilde{c} = c/c_{\max}$, $\mathbb{E}_0 = \lambda_0 \mathbf{1} \otimes \mathbf{1} + \mu_0 \mathbb{I}$ is the fourth-order isotropic elasticity tensor, where $\mathbf{1}$ and $\mathbb{I}$ are the unit second- and fourth- order tensors, respectively, $\lambda_0=v_0 E_0/[\left(1-2 v_0\right)\left(1+v_0\right)]$ and $\mu_0=E_0/[2\left(1+v_0\right)]$ are the Lam\'{e} constants of isotropic elasticity, expressed in terms of Young’s modulus $E_0$ and Poisson’s ratio $\nu_0$ of the material. The energetic degradation function $\omega (d)$ in the cohesive phase-field fracture model \cite{Wu2017JMPS} is expressed in terms of crack phase-field $d$ as following, 
	\begin{align}\label{eq:energy-degradation-function}
		\omega (d) 
		= \dfrac{(1-d)^2}{(1-d)^2 + a_{1} d ( 1- 0.5 d)}, \quad \text{with} \quad
		a_{1} = \dfrac{4 l_{\text{ch}}}{\pi b},
	\end{align}
	where $l_{\text{ch}} = \bar{E}_0 G(\boldsymbol{x}) / \sigma_c ^2$ is the Irwin’s internal length \cite{Wu2017JMPS,Chen2022TAFM}, and $\sigma_c$ is the failure strength (critical stress) under uni-axial tensile test, $\bar{E}_0 = E_0 (1-\nu_0) / (1+\nu_0) / (1-2\nu_0)$ expresses the longitudinal modulus.

	By inserting the variations of \cref{eq:potential-energy} into \cref{eq:energy-dissipation}, we can obtain
	\begin{align}\label{eq:energy-dissipation-further}
		\dot{\mathscr{D}} = 
		\int_{\varOmega} 
		\left( \bfsigma - \dfrac{\partial \psi}{\partial \bfepsilon} \right) 
			: \dot{\bfepsilon} \; \mathrm{d} V
		+ \int_{\varOmega} 
		\left( \mu - \dfrac{\partial \psi}{\partial c} \right) 
			: \dot{c} \; \mathrm{d} V
		- \int_{\varOmega} 
			\bfJ \cdot \nabla \mu \;
		\mathrm{d} V
		- \int_{\varOmega} 
		\left( \dfrac{\partial \psi}{\partial d} \dot{d} 
		+ \dfrac{\partial \psi}{\partial \nabla d} \cdot \nabla \dot{d} \right) 
		\mathrm{d} V
		\geq 0,
	\end{align}
	follow the Coleman-Noll principle \cite{Coleman1967} for satisfying the dissipation inequality, coupled chemo-mechanical constitutive equations can be obtained:
	\begin{subequations}\label{eq:constitutive-laws}
		\begin{align}
			\label{eq:constitutive-laws-mechanical}
			&
			\bfsigma = \dfrac{\partial \psi}{\partial \bfepsilon}
			= \dfrac{\partial \psi_m}{\partial \bfepsilon}
			= \omega (d) \mathbb{E}_0 : \left( \bfepsilon - \bfepsilon_c \right),
			\\
			\label{eq:constitutive-laws-chemical-potential}
			&
			\mu = \dfrac{\partial \psi}{\partial c}
			= \dfrac{\partial \psi_c}{\partial c} + \dfrac{\partial \psi_m}{\partial c}
			= R T \ln \dfrac{\tilde{c}}{1-\tilde{c}} - \bfsigma : \boldsymbol{\Omega},
			\\
			\label{eq:constitutive-laws-chemical-flux}
			&
			\bfJ = - \boldsymbol{M} (c, d) \cdot \nabla \mu
			= - \omega (d) \dfrac{c (1-\tilde{c})}{RT} \boldsymbol{D} \cdot \nabla \mu.
		\end{align}
	\end{subequations}
	In above formulations, $\boldsymbol{M} = \omega (d) c (1-\tilde{c})/(RT) \boldsymbol{D}$ is concentration-dependent mobility tensor, which is degraded by $\omega(d)$ in terms of crack phase-field; 
	% $\boldsymbol{D}$ denotes the diffusivity tensor with different diffusion speeds between the directions of in-plane (a-b) and out-of-plane (c) in the transversely isotropic model \cite{Jeff2021JPS,Peter2023ECActa}. 
	$\boldsymbol{D} = D_{ij} \boldsymbol{e}_{i} \otimes \boldsymbol{e}_{j}$ denotes the diffusivity tensor with $D_{ij}$ representing its components. 
	As can be seen from the thermodynamic derivations, chemical potential \cref{eq:constitutive-laws-chemical-potential} and chemical flux \cref{eq:constitutive-laws-chemical-flux} is affected by both the gradients of lithium concentration and mechanical stress. Under above settings of flux vector $\bfJ$ with linearly dependence on the gradient of chemical potential $\nabla \mu$, following term in \cref{eq:energy-dissipation-further} can be ensured non-negative, e.g.,
	\begin{align}
		- \int_{\varOmega} \bfJ \cdot \nabla \mu \; \mathrm{d} V
		= \int_{\varOmega} \omega (d) \dfrac{c (1-\tilde{c})}{RT} \boldsymbol{D} \cdot \nabla \mu \cdot \nabla \mu \; \mathrm{d} V \geq 0.
	\end{align}

	The remaining term in the dissipation inequality \cref{eq:energy-dissipation-further} after employing partial integration reads:
	\begin{align}\label{eq:energy-dissipation-damage}
		\int_{\varOmega} 
		\left[ -\omega'(d) \bar{Y} - \dfrac{G(\boldsymbol{x})}{b\pi} (2-2d) +
		\dfrac{2b}{\pi} G(\boldsymbol{x}) \nabla \cdot \nabla d
		\right] \dot{d} \;
		\mathrm{d} V +
		% surface for BC
		\int_{\partial \varOmega} - \dfrac{2b}{\pi} G(\boldsymbol{x}) \bfn \cdot \nabla d \; \dot{d} \; \mathrm{d} A \;
		\geq 0.
	\end{align}
	Call for the dissipative nature of damage evolution, i.e., $\dot{d} \geq 0$, by assuming the maximum dissipation principle and applying the Lagrange multipliers under Karush–Kuhn–Tucker (KKT) constraints, we obtain following governing equations for the crack phase-field,
	\begin{align}
		\label{eq:governing-equations-damage}
		\begin{cases}
			% case1
			Q + \nabla \cdot \boldsymbol{q} = 0,
			\quad \text{when } \dot{d} > 0 \vspace{1mm} 
			\quad \quad &
			\text{in} \; \varOmega\\
			% case2
			Q + \nabla \cdot \boldsymbol{q} < 0,
			\quad \text{when } \dot{d} = 0
			\quad \quad &
			\text{in} \; \varOmega \\
			% surface BC
			\nabla d \cdot \boldsymbol{n} = 0
			\quad \quad &
			\text{on} \; \partial \varOmega
		\end{cases},
		\quad \text{with} \quad
		\begin{cases}
			% case1
			Q = -\omega'(d) \bar{Y} - \dfrac{G(\boldsymbol{x})}{b\pi} (2-2d) \vspace{2mm} \\
			% case2
			\boldsymbol{q} = \dfrac{2b}{\pi} G(\boldsymbol{x}) \nabla d
		\end{cases},
	\end{align}
	with the crack driving force expressed as $\bar{Y} = \dfrac{1}{2} \left( \bfepsilon - \bfepsilon_c \right) : \mathbb{E}_0 : \left( \bfepsilon - \bfepsilon_c \right)$ from the thermodynamically consistent derivation.

	The aforementioned driving force does not distinguish between the asymmetric mechanical responses of solids under tension and compression. To prevent the fracture and damage to happen under compressive state, one may utilize either positive/negative \cite{Miehe2010IJNME} or volumetric/deviatoric \cite{AMM2009JMPS} decomposition of the strain, stress or effective stress, as extensively discussed in \cite{Van2020IJSS,Ziaei2023JMPS}. A simpler solution \cite{Wu2020CMAME,Chen2021CMAME} is to maintain isotropic stress-strain relation \cref{eq:constitutive-laws-mechanical} while consider contribution from positive effective stress in the mechanical driving force, i.e.,
	\begin{align}
		\label{eq:crack-driving-force}
		\bar{Y} = \dfrac{1}{2} \bfepsilon_e : \mathbb{E}_0 : \bfepsilon_e = 
		\dfrac{1}{2} \bar{\bfsigma} : \mathbb{S}_0 : \bar{\bfsigma}
		\qquad \quad \Longrightarrow \qquad \quad
		\bar{Y} = \dfrac{1}{2} \bar{\bfsigma}^+ : \mathbb{S}_0 : \bar{\bfsigma}^+ ,
	\end{align}
	where $\mathbb{S}_0$ is the compliance fourth-order tensor in mechanical sub-problem, and effective stress is defined as $\bar{\bfsigma} = \mathbb{E}_0 : \bfepsilon_e = \mathbb{E}_0 : \left( \bfepsilon - \bfepsilon_c \right)$. In this work, the positive/negative projection of the effective stress in energy norm \cite{Wu2020CMAME} is adopted, with the positive cone expressed as $\overline{\boldsymbol{\sigma}}^{+}=\sum_{i=1}^3 \bar{\sigma}_i^{+} \boldsymbol{v}_i \otimes \boldsymbol{v}_i$, for the $i$-th principal value $\bar{\sigma}_i^{+}$ and the corresponding principal vector $\boldsymbol{v}_i$ of the effective stress tensor $\bar{\bfsigma}$. For simplicity, this work exclusively considers the simplest case, wherein crack evolution is driven solely by its major principal value \cite{Wu2020CMAME,Chen2021CMAME}, thus,
	\begin{align}
		\label{eq:crack-driving-force-simple}
		\bar{Y} = \frac{ \bar{\sigma}_{\text{eq}}^2 }{2 \bar{E}_0}, 
		\qquad \text{with} \quad
		\bar{\sigma}_{\text{eq}} = \max( \left\langle \bar{\sigma}_1 \right\rangle, \sigma_c).
	\end{align}
	In the above, the Macaulay brackets $\left\langle \cdot \right\rangle$ are defined as $\left\langle x \right\rangle = \max (x, 0)$, and $\sigma_c$ means uni-axial tensile strength as mentioned before. By redefining the crack driving force $\bar{Y}$, it becomes feasible to effectively capture tension-dominant fracture behaviors in cathode particles. It's worth noting that while such modifications and the aforementioned constitutive relations may not be variationally consistent, they do uphold thermodynamic consistency in terms of energy dissipation \cite{Wu2020CMAME}.

	\subsection{Summary of chemo-mechanical cohesive phase-field fracture model}
	The governing equations, constitutive equations and boundary conditions of chemo-mechanical cohesive phase-field fracture model are summarized in \cref{fig:configuration} (b), wherein the displacement field $\bfu(\boldsymbol{x})$, lithium concentration field $c(\boldsymbol{x})$, and crack phase-field $d(\boldsymbol{x})$ serve as primary unknowns in the mechanical, chemical, and fracture sub-problems, respectively.
	
	The couplings of different physics are also shown in \cref{fig:configuration} (b), as can be seen, chemical and mechanical processes are fully coupled, i.e., changing lithium concentration leads to chemical deformation and mechanical stress, which conversely influences the distribution of lithium concentration. Meanwhile, mechanical and fracture processes exhibit a two-way coupled relationship, e.g., mechanical stress drives the evolution of cracks and fractures, which conversely damage the structural integrity and degrade its bearing capacity. Additionally, the fracture process exerts an influence on the chemical sub-problem, specifically, through the degradation of diffusivity with the presence of crack, which obstructs the transport of lithium; while the direct effect of chemical process on fracture sub-problem is not integrated  into the current framework, which can be extended in our future work by considering concentration-dependent fracture property \cite{Shahed2023TAFM, Jeff2021JPS} or cycles induced materials' embrittlement \cite{Emilio2022JPS}.

	% ***********************************************
	\section{Numerical examples and results}
	\label{results}
	
	In this section, a series of numerical examples regarding bulk (trans-granular) and interface (inter-granular) fractures in purely-mechanical and chemo-mechanical problems are simulated using the proposed model. Firstly, a square-shape specimen with a predefined horizontal notch and an internal interface (grain boundary) under tensile loading is presented as purely-mechanical benchmark. Secondly, for validating the applicability to chemo-mechanically coupled scenario, similar specimens under chemical delithiation mimicking the charging condition of cathode particles are considered. For aforementioned two cases, the dependence of numerically predicted results on phase-field and interface length-scale parameters ($b$ and $L$) within various interface models are studied. Moreover, predictions given by cohesive phase-field fracture model are compared and validated with predictions by the cohesive zone model (see \cref{appendix-chemo-mechanical-czm} for chemo-mechanical fracture formulations). Eventually, the proposed model is applied to image-based reconstructed 3D polycrystalline geometry, simulating its diverse (inter-/trans- granular) fracture patterns in NMC cathode particles.
	
	The cohesive phase-field fracture models (under both purely-mechanical and chemo-mechanical circumstances) are implemented into the open-source FEM framework Multiphysics Object-Oriented Simulation Environment (MOOSE, \texttt{https://mooseframework.inl.gov/index.html}). The geometry and finite-element meshes of benchmark examples are generated with the aid of the open-source software Gmsh (\texttt{https://gmsh.info/}). Finally, the visualization of obtained results are carried out on ParaView (\texttt{https://www.paraview.org/}). The utilization of energy degradation function in \cref{eq:energy-degradation-function} in the CPF model necessitates the fulfillment of its convexity w.r.t. phase-field variable, thereby imposing specific criteria on the value of phase-field length-scale $b$. Besides, the element size $h$ and the phase-field length-scale $b$ satisfy $h \leq b/3$, in order to resolve the gradient of the crack phase-field. Interested readers can refer to \cite{Chen2022TAFM,Chen2024JPS} for details.

	\subsection{Purely mechanical benchmark: Notched plate under tensile loading}
\begin{figure}[htbp]
		\centering
		\includegraphics[width=0.95\textwidth]{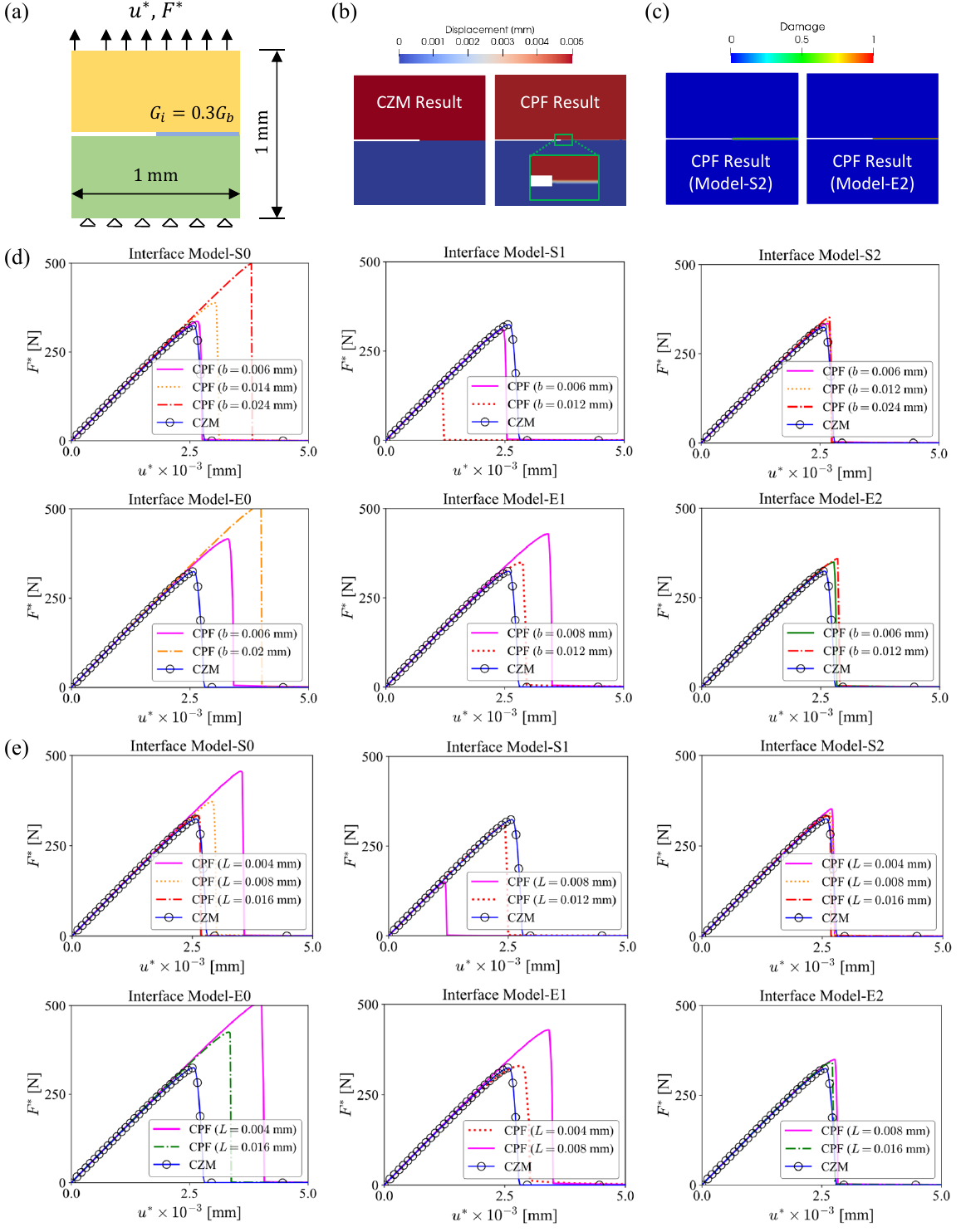}
		\caption{Single-edge notched plate with a horizontal interface under vertical mechanical loading. Figure (a) schematically illustrates the geometry with predefined notch and GB, and also the boundary conditions. Figure (b) compares the displacement results obtained by the CZM and the CPF model. Figure (c) compares the crack results obtained by phase-field interface fracture Model-S2 and Model-E2. Figure (d) compares structural reaction forces predicted by the CPF model and CZM in order to study the influence of $b$; for all simulations conducted with CPF, identical value $L = 0.008 $ mm is adopted. Figure (e) compares structural reaction forces predicted by the CPF model and CZM for studying the influence of $L$, identical value $b = 0.012 $ mm is adopted for all simulations conducted with CPF.}
		\label{fig:example1}  
	\end{figure}

	Let us first consider one benchmark problem under external mechanical loadings, in order to validate capability of the proposed CPF model for simulating interface fracture. As shown in \cref{fig:example1} (a), it is a square plate of length 1 mm, with unit out-of-plane thickness. A straight horizontal notch, measuring 0.5 mm in length, is introduced at the half height of the specimen; and a horizontal internal interface (grain boundary) connects between the right end of the notch and the midpoint of specimen's right edge, bridging the upper and lower grains. The bottom edge is fixed, while a vertical displacement is applied on the top edge.
	
	The purely mechanical CPF model can be obtained by switching off the chemical process in the fully coupled model in \cref{sec:phase-field}. The corresponding material parameters adopted in the simulations are listed in \cref{table:material-parameters-mechanical}, here the grain boundary possesses relatively weaker strength and fracture energy than the grains. In order to study the influences of length-scale parameters on numerical results, parametric studies with diverse values of $b$ and $L$ within different interface models are investigated. Besides, for numerical simulations conducted with the cohesive zone model, identical values in \cref{table:material-parameters-mechanical} are used, except two length-scale parameters which are not involved in the CZM.
	
	\begin{table}[htbp]
		\centering
		\caption{Material parameters in the purely-mechanical example taken from \cite{Paggi2017CMAME, Wu2017JMPS, Hansen2019CMAME}}
		\label{table:material-parameters-mechanical}
		\begin{tabular}{ m{0pt} m{7cm}<{\centering} m{2cm}<{\centering} m{2.5cm}<{\centering} }
			\toprule[1pt]
			\rule{0pt}{10pt} &
			Material parameters & Value & Unit \\
			\hline
			%
			%\hline
			\rule{0pt}{10pt} 
			& Young’s modulus $E_{0}$ 
			& 210 
			& [GPa] \\
			% 
			%\hline
			\rule{0pt}{10pt} 
			& Poisson's ratio $\nu_{0}$ 
			& 0.3 
			& [-] \\
			% 
			%\hline
			\rule{0pt}{10pt} 
			& Tensile strength of grain $\sigma_{c,b}$ 
			& 2400
			& [MPa] \\
			% 
			%\hline
			\rule{0pt}{10pt} 
			& Tensile strength of grain boundary $\sigma_{c,i}$ 
			& 1315
			& [MPa] \\
			% 
			%\hline
			\rule{0pt}{10pt} 
			& Fracture energy of grain $G_{b}$ 
			& 2700 
			& [N/m] \\
			%\hline
			\rule{0pt}{10pt} 
			& Fracture energy of grain boundary $G_{i}$ 
			& 810 
			& [N/m] \\
			% \hline
			\rule{0pt}{10pt} 
			& Phase-field length-scale parameter $b$ 
			& 0.006 $\sim$ 0.024
			& [mm] \\
			% 
			% \hline
			\rule{0pt}{10pt} 
			& Interface length-scale parameter $L$ 
			& 0.004 $\sim$ 0.016
			% & [$\upmu$m] \\
			& [mm] \\
			\bottomrule[1pt]
		\end{tabular}
	\end{table}

	Along with increasing mechanical loading, the crack initially originates at the tip of the notch due to local high stress level, then propagates along the horizontal interface towards the right edge of the specimen, ultimately  resulting in the structural failure. \cref{fig:example1} (b) compares the displacement results at the moment when crack fully propagates (see \cref{fig:example1} (c)), which are obtained by the CPF model and the CZM, respectively. A distinct displacement jump is observable in the CZM at the grain boundary where interface fracture occurs, while the displacement field in the phase-field simulation maintains continuity with a smooth transition at the crack site. Despite these differences, both models yield consistent predictions regarding the global displacement profile. Besides, interface Model-S2 and Model-E2 give consistent predictions on fracture evolution path, despite different (stair-wise of exponential) representations of interface adopted in the simulations.

	What draws more attentions are quantitative structural reaction forces in \cref{fig:example1} (d) and (e). Numerically predicted loads $F^{*}$ versus displacements $u^{*}$ under varying $b$ and $L$ are compared. The results produced by the CPF model, reveal consistent trends with an initially linear elastic regime followed by a sudden drop attributed to the complete propagation of brittle cracks. Nevertheless, interface Model-S0, Model-S1, Model-E0 and Model-E1 exhibit inconsistent peak structure bearing capacities with predictions from the cohesive zone model, demonstrating significant dependencies on length-scale parameters $b$ and $L$. This can be attributed to the imprecise evaluation of interface fracture energy, as listed in \cref{table:interface-models}. With introduced effective interface fracture energy in the CPF framework to realize the equivalence to the sharp interface fracture energy, length-scale insensitive results consistent with CZM predictions can be attained, irrespective of whether stair-wise (Model-S2) or exponential (Model-E2) representation of interface fracture energy is employed. 

%\vspace{-10mm}
	
	\subsection{Chemo-mechanical benchmarks: Notched plates under de-lithiation}
	
	We are now looking into the chemo-mechanical case where the lithium flow drives the cracking progression in the charging (de-lithiation) process. The geometry adopted in the simulation is similar to the purely-mechanical case, specifically, a square-shape specimen consisting of 2-grains with connected grain boundaries and an introduced notch defect is considered. Two cases with notches of varying lengths are presented. 
	\begin{table}[htbp]
		\centering
		\caption{Material parameters in the chemo-mechanical examples taken from \cite{Jeff2021JPS,Avtar2022JMPS,Shahed2021JMPS,Bai2021IJSS,Xu2018JMPS}}
		\label{table:material-parameters-chemo-mechanical}
		\begin{tabular}{ m{0pt} m{7cm}<{\centering} m{2cm}<{\centering} m{2.5cm}<{\centering} }
			\toprule[1pt]
			\rule{0pt}{10pt} &
			Material parameters & Value & Unit \\
			\hline
			%
			%\hline
			\rule{0pt}{10pt} 
			& Young’s modulus $E_{0}$ 
			& 93000 
			& [GPa] \\
			% 
			%\hline
			\rule{0pt}{10pt} 
			& Poisson's ratio $\nu_{0}$ 
			& 0.3 
			& [-] \\
			% 
			%\hline
			\rule{0pt}{10pt} 
			& Tensile strength of grain $\sigma_{c,b}$ 
			& 600
			& [MPa] \\
			% 
			%\hline
			\rule{0pt}{10pt} 
			& Tensile strength of grain boundary $\sigma_{c,i}$ 
			& 425
			& [MPa] \\
			%\hline
			\rule{0pt}{10pt} 
			& Fracture energy of grain $G_{b}$ 
			& 5 
			& [N/m] \\
			% 
			%\hline
			\rule{0pt}{10pt} 
			& Fracture energy of grain boundary $G_{i}$ 
			& 2.5 
			& [N/m] \\
			%\hline
			\rule{0pt}{10pt} 
			& Maximum Lithium concentration $c_{\max}$ 
			& 22900
			& [mol/m$^3$] \\
			% 
			%\hline
			\rule{0pt}{10pt} 
			& Partial molar volume $\Omega$ 
			& 3.497 $\times$ 10$^{-6}$
			& [m$^3$/mol] \\
			% 
			%\hline
			\rule{0pt}{10pt} 
			& Diffusivity $D$ 
			& 7 $\times$ 10$^{-15}$
			& [m$^2$/s] \\
			% 
			%\hline
			\rule{0pt}{10pt} 
			& Gas constant $R$ 
			& 8.314
			& [J/(mol$\cdot$K)] \\
			% 
			%\hline
			\rule{0pt}{10pt} 
			& Temperature $T$ 
			& 298.15
			& [K] \\
			% \hline
			\rule{0pt}{10pt} 
			& Phase-field length-scale $b$ 
			& 0.06 $\sim$ 0.24
			& [$\upmu$m] \\
			% 
			% \hline
			\rule{0pt}{10pt} 
			& Interface length-scale $L$ 
			& 0.04 $\sim$ 0.16
			% & [$\upmu$m] \\
			& [$\upmu$m] \\
			\bottomrule[1pt]
		\end{tabular}
	\end{table}

	\subsubsection{Case 1: Inter-granular fracture of the specimen}

\begin{figure}[htbp]
		\centering
		\includegraphics[width=0.95\textwidth]{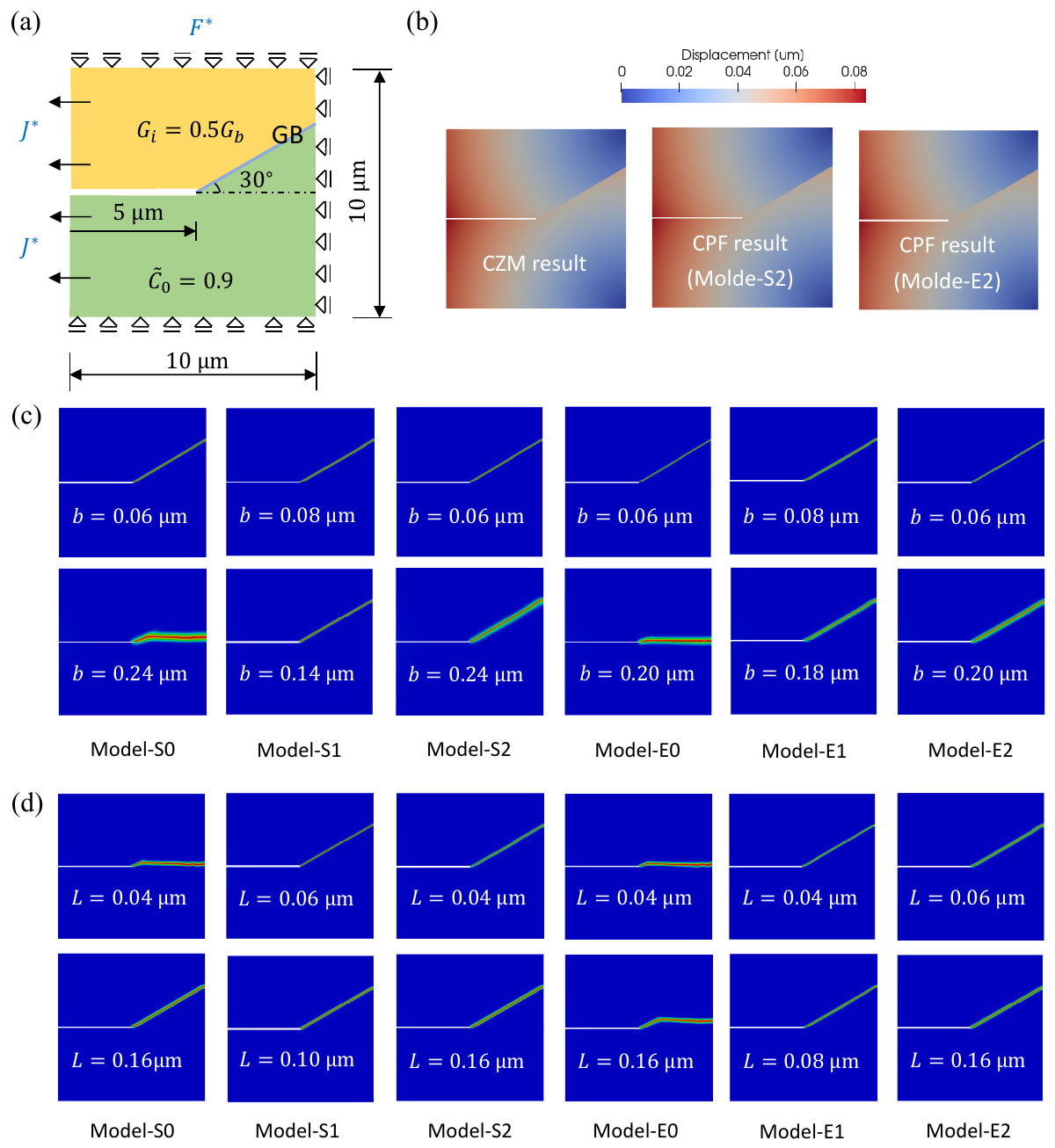}
		\caption{Single-edge notched plate with a 30-degrees slope GB under de-lithiation induced chemical flux. Figure (a) schematically illustrates the geometry, grain boundary, notch, mechanical and chemical boundary conditions. Figure (b) compares displacement results obtained by the CZM and the CPF model. Figure (c) and (d) compare crack phase-field results obtained by various CPF interface models under varying length-scale parameters $b$ and $L$.}
		\label{fig:example2-1}
	\end{figure}

	In order to study inter-granular (interfacial) chemo-mechanical fracture and the corresponding structural responses, as shown in \cref{fig:example2-1} (a), the notch is assumed to be half long of the specimen, so that crack can nucleate and propagate in the interface. A sharp grain boundary of 30$^{\circ}$ slope with relatively weaker mechanical failure resistance than grain is introduced. The whole system has initial concentration ($\tilde{c}_0 = c_0 / c_{\max} = 0.9$) expressed in a normalized fashion and is assumed to be the stress-free initial state. Properly prescribed chemical and mechanical boundary conditions are considered, i.e., the de-lithiation flux is applied at the left edge and the displacement components along the normal directions of the left three edges are constrained.

	The material parameters and their numerical values in the modeling are summarized in \cref{table:material-parameters-chemo-mechanical}. Sharp GB is assumed to have the fracture energy $G_i$ equaling to half that of the grains ($G_b$). Here the diffusivity (in \cref{eq:constitutive-laws-chemical-flux}) and partial molar volume (in \cref{eq:kinematics-chemical}) are assumed to be isotropic in the benchmark simulations for simplicity, i.e., $D_{ij} = D \delta_{ij}$ and $\Omega_{ij} = \Omega \delta_{ij}$ with $\delta_{ij}$ denoting the Kronecker Delta Function. A range of length-scale parameters ($b$ and $L$) are considered to investigate their respective influences on numerical predictions of chemo-mechanical fracture behaviors.

 \begin{figure}[htbp]
     \centering
     \includegraphics[width=0.95\textwidth]{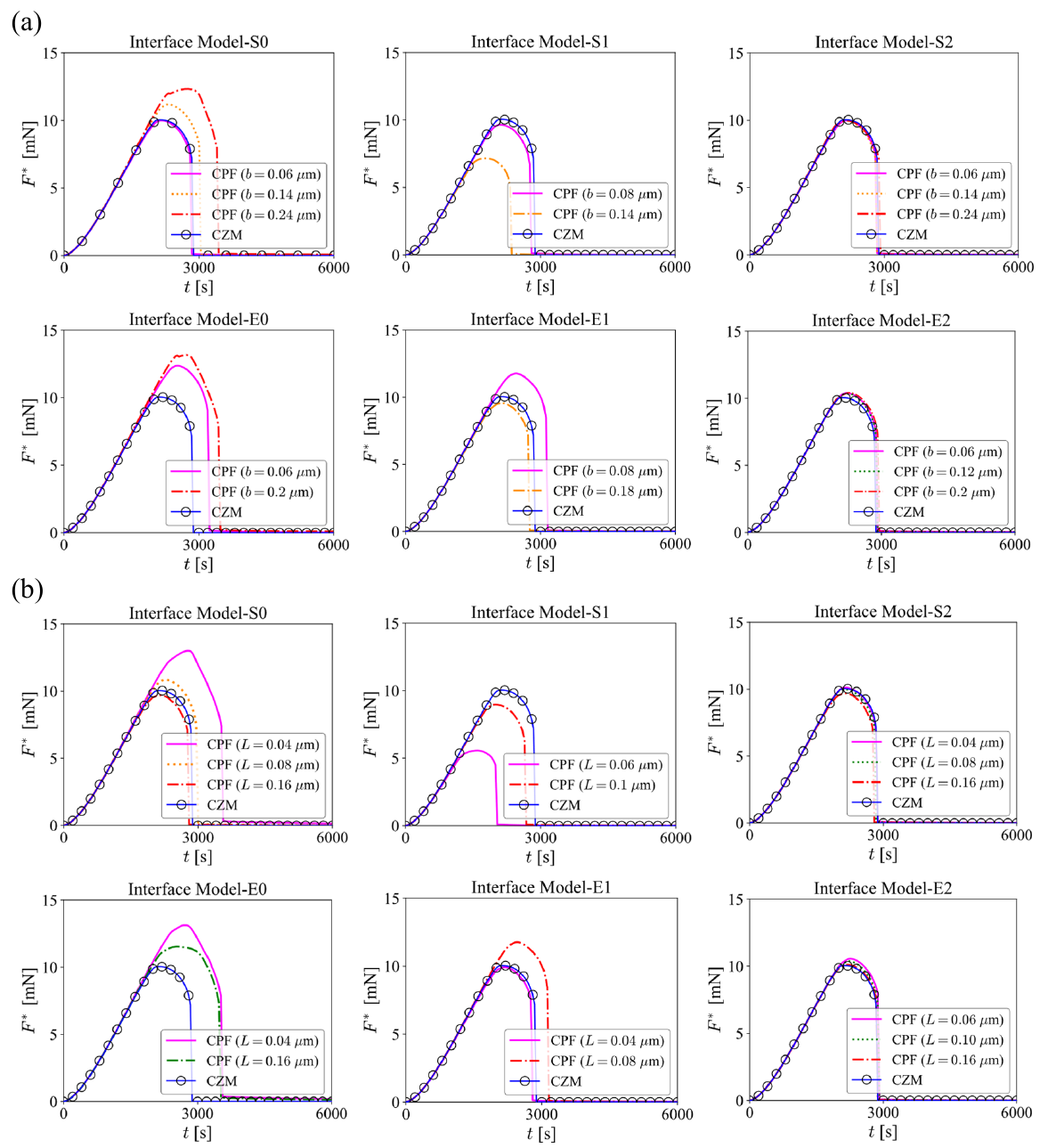}
     \caption{Figure (a) and (b) compare the structural reaction forces predicted by the CPF model and the CZM to study the influence of length-scale parameters $b$ and $L$. Note for CPF simulations, a fixed value $L = 0.08 $ $\upmu$m is adopted for studying the influence of $b$, while $b = 0.12$ $\upmu$m is fixed when investigating the effect of $L$.}
     \label{fig:example2-1-b}
 \end{figure}

	As the lithium goes out from the left edge, the specimen undergoes volume shrinkage. Due to the mechanical boundary conditions on the other three edges constraining the deformation, tensile stress develops within the structure and crack nucleates at the tip of the introduced notch. As more delithiation flux takes place, the crack propagates further along the weaker grain boundary till the specimen is completely divided into two separate parts. \cref{fig:example2-1} (b) compares the displacement profiles obtained by the CPF model and the CZM, after the complete propagation of inter-granular crack. Cohesive zone model and phase-field model give consistent predictions on global mechanical deformations of specimen, with observable displacement jumps at the grain boundary where fracture occurs. \cref{fig:example2-1} (c) and (d) compare predicted crack patterns under varied choices of parameters $b$ and $L$. Despite different values of length-scale parameters adopted in the simulations, Model-S2 and Model-E2 yield consistent predictions of fracture patterns, by employing the precisely determined effective interface fracture energy so that the interface fracture energy can be ensured to be equivalent to the sharp case. However, other interface models such as Model-S0 and Model-E0 obtain inconsistent predictions on failure patterns, which are length-scale sensitive/dependent, e.g., a larger $b$ or a smaller $L$ in those models tends to significantly overestimate the mechanical failure resistance of the interface, so that crack pattern is observed to deflect from the interface and into the grain; while crack is predicted to propagate along the GB when $b$ is smaller or $L$ is larger.

	Same conclusion can also be drawn from the comparison of quantitative structural reaction-force curves in \cref{fig:example2-1-b} (a) and (b). As can be seen, the overall global responses in Model-S2 and Model-E2, which agree well with predictions from the CZM, are negligibly affected by the choices of internal length-scale parameters $b$ and $L$. In contrast, due to incorrect assess of fracture energy at the interface by Model-S0, Model-S1, Model-E0 and Model-E1, unconvincing results are obtained, which show strong dependency/sensitivity on the choices of length-scale parameters. 

	\subsubsection{Case 2: Inter- and trans-granular fractures of the specimen}

\begin{figure}[htbp]
		\centering
		\includegraphics[width=0.9\textwidth]{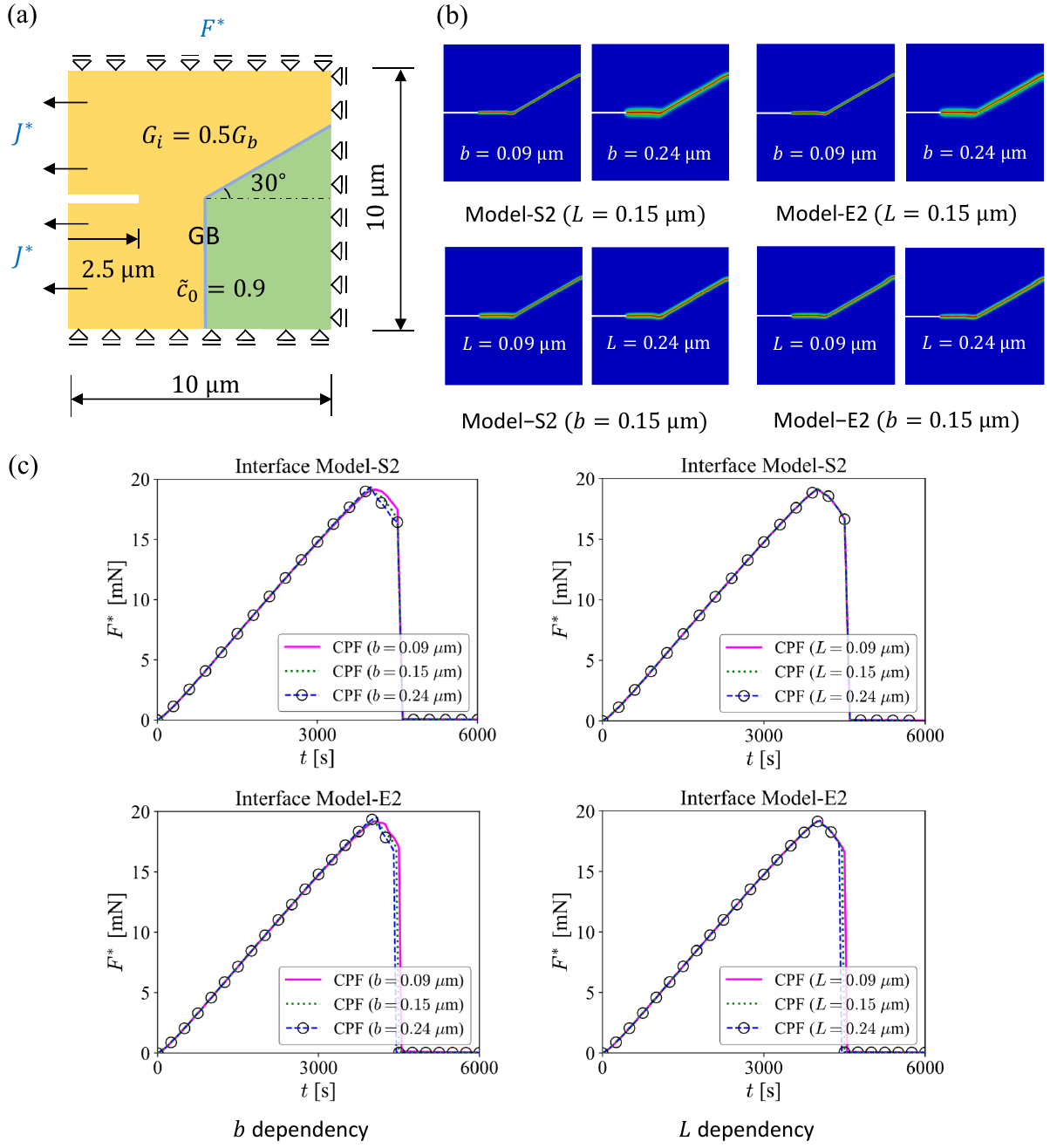}
		\caption{Single-edge notched plate with grain boundary and notch under de-lithiation induced chemical flux and proper mechanical constraints. Figure (a) schematically illustrates the geometry, grain boundary, mechanical and chemical boundary conditions. Figure (b) compares the phase-field crack results obtained by CPF Interface Model-S2 and Model-E2 with varying length-scale parameters $b$ and $L$. Figure (c) shows the structural reaction forces for illustrating the independence of obtained results on length-scale parameters $b$ and $L$.}
		\label{fig:example2-2}
	\end{figure}

	In this subsection, we employ the proposed chemo-mechanical CPF model to analyze an example encompassing both inter- and trans-granular fractures. This serves the purpose of simulating potential crack paths in polycrystalline cathodes, which initially start in the grain and then evolve towards grain boundaries. As shown in \cref{fig:example2-2} (a), two individual grains are connected with the grain boundary. One horizontal notch with quarter length of the specimen is introduced in the left grain, so that trans-granular crack could nucleate here. The simulation setups, e.g., materials properties, initial and boundary conditions, can refer to the previous example. Given the limitations of the CZM in modeling bulk fracture or trans-granular cracking behaviors \cite{CZM2002}, in this example, only phase-field modeling is considered, specifically, Model-S2 and Model-E2 are adopted in the simulations to validate their respective performances. 

	\cref{fig:example2-2} (b) compares the fracture patterns obtained with Model-S2 and Model-E2. As the lithium flux goes out from the left edge, chemical deformation triggers local high mechanical stress at the notch tip, initiating crack nucleation and propagation. Over time and as delithiation progresses, the fracture propagates initially within the left grain and subsequently deflects towards the GB with relatively weaker failure resistance, leading to intricate inter- and trans-granular failure modes. As can be seen, identical crack patterns are obtained in the simulations in spite of varied choices of length-scale parameters $b$ and $L$, which only affects the width of the diffusive phase-field damage band. Regardless of stair-wise (in Model-S2) or exponential (in Model-E2) representation of interface in the modeling, the approach ensuring interface fracture energy consistent with that in the cohesive zone model is very robust in chemo-mechanical phase-field fracture simulations, which gives length-scale insensitive numerical predictions.

	Furthermore, \cref{fig:example2-2} (c) compares reaction force w.r.t. delithiation time obtained with Model-S2 and Model-E2. As can be concluded, internal length-scale parameters $b$ and $L$ have negligible effects on global structural responses, which applies to both stair-wise (Model-S2) and exponential (Model-E2) interpolations of fracture energy.

	\subsection{Chemo-mechanical application: Image-based reconstructed 3D NMC cathode geometry}
	
\begin{figure}[htbp]
		\centering
		\includegraphics[width=0.85\textwidth]{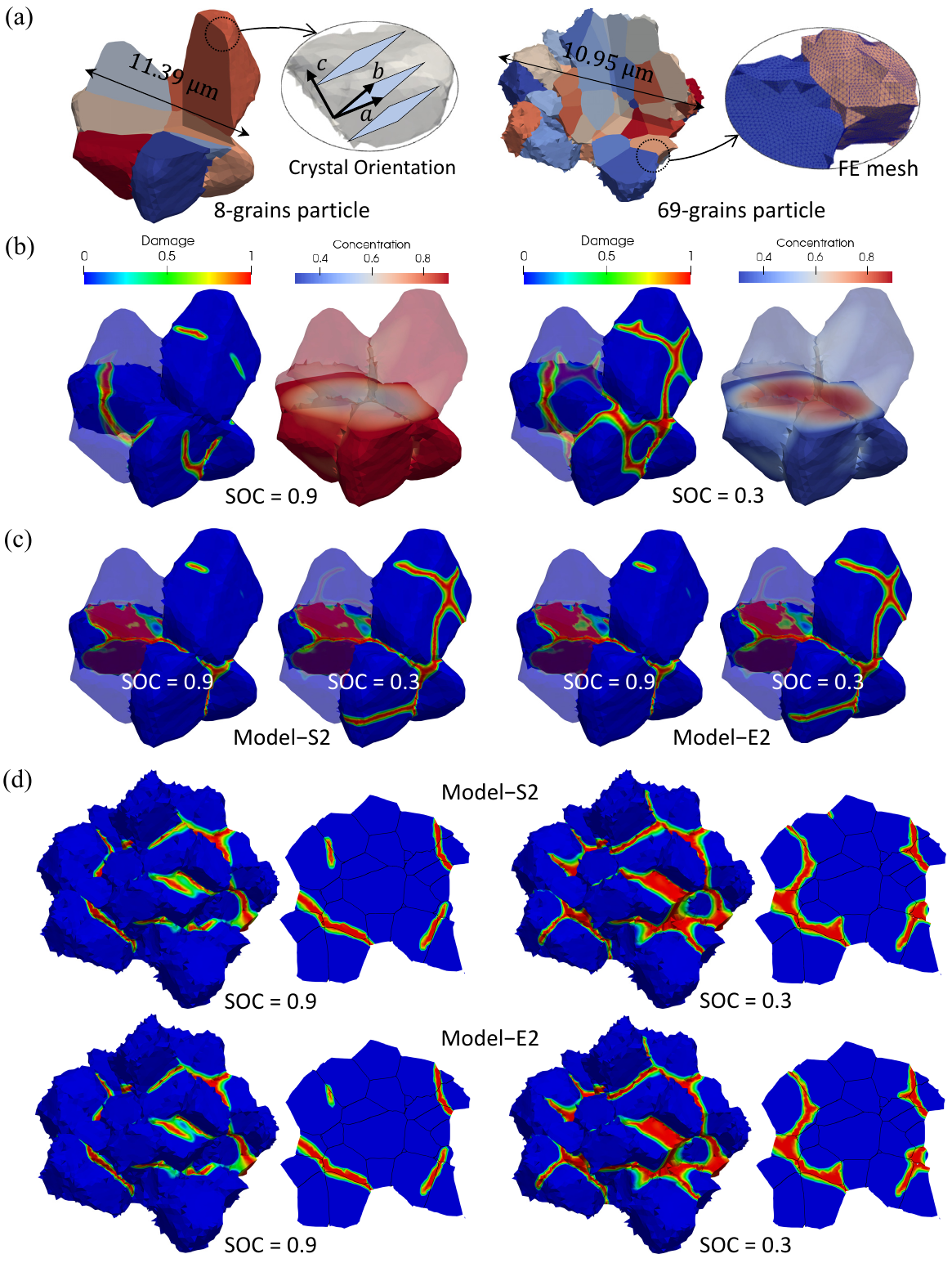}
		\caption{Chemo-mechanical simulation results on image-based reconstructed 3D NMC cathode particles. Figure (a) shows the geometry and polycrystalline microstructure of two particles consisting of 8 and 69 grains, respectively; finite-element meshes and crystal orientations for transversely isotropic model are also illustrated. Figure (b) shows the concentration and crack phase-field results at the ending moments of lithiation (SOC=0.9) and de-lithiation (SOC=0.3), respectively. Figure (c) and (d) compare the fracture patterns of particles comprising of 8 and 69 grains, respectively, which are obtained by two interface models (Model-S2 and Model-E2). For the 69-grains particle case, fracture results within a section are also depicted.}
		\label{fig:example3}
	\end{figure}

	In the present section, the proposed chemo-mechanical phase-field model is applied to image-based reconstructed 3D NMC cathode particles, see \cref{fig:example3} (a) for the visualization of the geometry. In addition to ensuring consistency of interface fracture energy with the cohesive zone model and providing length-scale insensitive predictions, the proposed new Model-E2 is particularly notable for its flexibility in handling intricate grain boundaries/interface topologies within 3D polycrystalline microstructures, which will be highlighted in this example.

	The generation and meshing of cathode polycrystalline geometry are briefly reviewed, readers can refer to \cite{Furat2021, Chen2024JPS} for further details on this process. Statistical characterizations of the outer shell of Li$_x$Ni$_{0.5}$Mn$_{0.3}$Co$_{0.2}$O$_2$ (NMC532) particles are obtained from nano-computed tomography (nano-CT) data, while grain architectures are determined from focused-ion beam (FIB) and electron backscatter diffraction (EBSD) data, integrated within a 3D stochastic model. Subsequently, two models are combined to generate virtual NMC particles together with their inner grain architectures, which possess statistically similarities to those observed in the nano-CT and FIB-EBSD data \cite{Furat2021}. In phase-field simulations, a subset of 8 and 69 grains composing the secondary particle with a volume of 370 $\upmu \text{m}^3$ are extracted from the full-grain architecture \cite{Jeff2021JPS}, which are eventually processed with \texttt{Iso2Mesh} (an open-source MATLAB code that converts images into meshes \cite{PYF20}) for finite-element mesh generations, as visualized in \cref{fig:example3} (a).

	\begin{figure}[htbp]
		\centering
		\includegraphics[width=\textwidth]{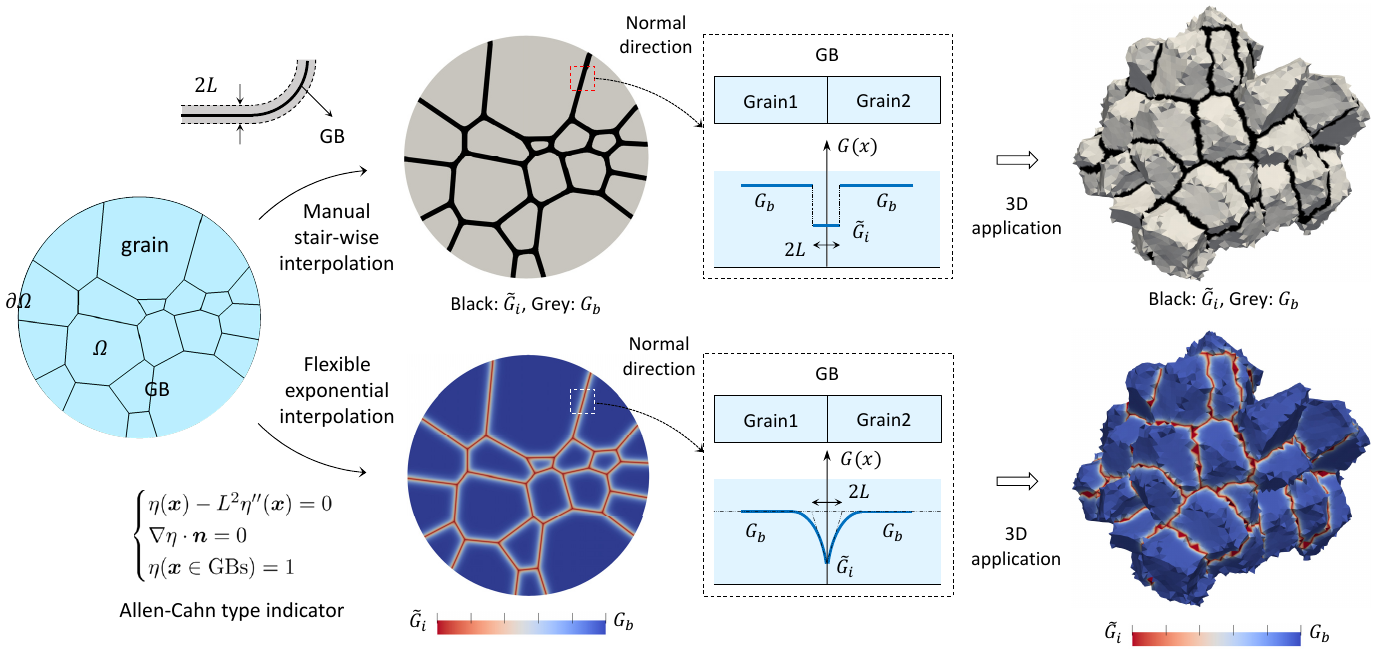}
		\caption{Comparison of implementations of stair-wise (in Model-S2) and exponential (in Model-E2) functions for the interpolation of fracture energy between the grains and the GBs in a polycrystalline microstructure.}
		\label{fig:example3-stairwise-exponential}
	\end{figure}

	To assign the fracture energy, i.e., $G(x)$ in \cref{eq:G(x)}, within the polycrystalline domain for simulations, implementations of stair-wise and exponential functions for interpolating the fracture energy between the grains and the grain boundaries are illustrated in \cref{fig:example3-stairwise-exponential}. As shown for Model-S2, a diffusive interface region with a width of $2L$ in the normal direction of the GBs needs to be manually generated, where the effective interface fracture energy $\tilde{G}_i$ is assigned; while the left regions possess the bulk fracture property $G_b$, so that stair-wise distribution $G(x)$ can be observed in the zoomed call-out. However, when dealing with a 3D polycrystalline structure with complex interface topologies, manually selecting the material points within the $2L$-wide diffusive interface region becomes increasingly cumbersome to implement. Instead, Model-E2 employs an Allen-Cahn type GB (interface) indicator $\eta(\bfx)$ to achieve the exponential interpolation of fracture energy between GBs and grains. This approach offers greater flexibility in handling intricate GB topologies, as demonstrated in 2D and 3D polycrystalline microstructure, it allows a smooth transition of fracture property, eliminating the need to manually define the diffusive GB region with a prescribed width, as required in Model-S2.

	In the modeling, material properties can refer to \cref{table:material-parameters-chemo-mechanical} of the previous example. In order to consider anisotropic behaviors in reconstructed 3D particles, a transversely isotropic model \cite{Jeff2021JPS,Peter2023ECActa} distinguishing properties between the directions of in-plane (a-b) and out-of-plane (c) is considered, see \cref{fig:example3} (a) for an illustration; specifically, for the diffusivity tensor and partial molar volume tensor,
	\begin{align}
		\boldsymbol{D} = 
			\boldsymbol{R} \left[
				\begin{array}{ccc}
					D_{ab} &0 &0  \\
					0 &D_{ab} &0  \\
					0 &0 &D_{c}
				\end{array}
			\right] \boldsymbol{R}^{\text{T}}
		, \qquad
		\boldsymbol{\boldsymbol{\Omega}} = 
			\boldsymbol{R} \left[
				\begin{array}{ccc}
					\Omega_{ab} &0 &0  \\
					0 &\Omega_{ab} &0  \\
					0 &0 &\Omega_{c}
				\end{array}
			\right] \boldsymbol{R}^{\text{T}},
	\end{align}
	here $\boldsymbol{R}$ denotes the rotation matrix determined from the crystal orientation of the grain. According to literature \cite{Jeff2021JPS,yamakawa2014effect,roberts2014framework,taghikhani2023electro,lim2018fundamental}, the in-plane diffusion coefficient is typically assumed to be 100 times faster as compared to the out-of-plane one, i.e., $D_{ab} = 100 D_{c} = D$; similarly, the partial molar volume is assumed to be 5 times larger in the out-of-plane direction as compared to the in-plane, namely, $\Omega_{ab} = 5 \Omega_{c} = \Omega$. It has been observed from several previous benchmarks that the length-scale parameters have negligible influence on numerical predictions, such as crack evolution paths and quantitative global responses. Therefore, for the simulations in this subsection, we consider $b = 0.4 \; \upmu \text{m}$ and $L = 0.2 \; \upmu \text{m}$.
	
	Regarding the boundary conditions, rigid-motion suppression is specified by fully constraining translation on one central point and partially constraining rotation on two points on the outside surface \cite{Jeff2021JPS, Chen2024JPS}. The electrostatic and electrochemical potential at the interface between the active material and the electrolyte determines how chemical reaction (Li$^+ + e^- \rightleftharpoons$ Li) takes place and the lithium flux on the secondary-particle surface, which can be described by the modified Butler–Volmer (BV) equation \cite{Zhao2016CMAME,Avtar2022JMPS,Shahed2021JMPS},
	\begin{align}
		\label{eq:BV-flux}
		J^* = \frac{c_{\text {surf }}}{\tau_0} (1-\tilde{c})
		\left[ 
		\exp \left(-\frac{F}{2 R T} \Delta \phi \right)
		- \exp \left( \frac{\mu}{R T} + \frac{F}{2 R T} \Delta \phi \right)
		\right],
	\end{align}
	where $c_{\text {surf}}$ is the molar concentration of intercalation sites on the surface, $\tau_0$ denotes the mean duration for a single reaction step which accounts for the slow and fast reaction, $F$ is Faraday's constant, $\Delta \phi = \phi_{\text{e}} - \phi(\tilde{c})$ is the voltage difference across the interface (refer to \cite{colclasure2019requirements,Jeff2021JPS} for detailed information) and $\mu$ expressed in \cref{eq:constitutive-laws-chemical-potential} is the chemical potential at the interface between active material and the electrolyte. 
	% The above normal flux in \cref{eq:BV-flux} can be directly used as a boundary condition to simulate charging/discharging scheme.
	For the given 10C-rate used in the simulation, one circle with initially discharging (lithiation) from $\text{SOC} = 0.3$ to 0.9 and subsequently charging (de-lithiation) till $\text{SOC} = 0.3$ is adopted as chemical boundary condition to simulate charging/discharging scheme.

	The influence of sharp interface fracture energy on crack patterns is firstly studied. \cref{fig:example3} (b) illustrates the concentration and crack phase-field results, under the assumption that grain boundaries and grains have identical fracture energy ($G_i = G_b$). As can be seen, during the first half stage with increasing lithium concentration and SOC (state of charging, to denote the average Li concentration in units of $c_{\max}$), lithium insertion induces volume expansion and correspondingly mechanical stress leads to crack nucleation and evolution from the interior of the structure to the exterior. During de-lithiation, as the lithium concentration decreases, tensile stress develops on the exterior of the particle due to global volume shrinkage. This stress leads to further propagation of cracks, eventually splitting the grains. Besides, intricate patterns such as branching and merging of cracks can be modeled by phase-field method under given fast charging scheme. For the case when grain boundaries have lower fracture energy than the grains ($G_i = 0.5 G_b$), as shown in \cref{fig:example3} (c), during the lithiation stage, cracks tend to emerge and grow at the GBs with lower mechanical failure resistance, as a result, an inter-granular crack mode can be observed; under de-lithiation condition, trans-granular fractures occur within the grains and connect with the inter-granular ones, with some exhibiting branching patterns.

	Subsequently, the influence of grain size on fracture behaviors of polycrystalline cathode particles is investigated with the proposed model. \cref{fig:example3} (d) illustrates the crack phase-field results of a reconstructed 3D cathode particle, which has the same volume as the 8-grains particle but contains more (69) smaller grains. As can be observed, during the lithiation stage, cracks nucleate from the interior and evolve to the exterior; after switching to de-lithiation condition, more cracks appear in the structure with some merging together, leading to global degradation of the cathode particle. Comparing the fracture behaviors of particles with the same volume but different grain size, it can be concluded that, inter-granular cracks at the GBs predominantly dictate the failure mode when grain size is smaller, while both inter- and trans-granular patterns can happen when grain size is larger, which agrees with the findings in our previous study \cite{Chen2024JPS}.

	% Furthermore, the phase-field cracks obtained by using interface Model-S2 and Model-E2 exhibit perfect agreement, thereby validating the importance of achieving interface fracture energy equivalence to a sharp crack, which ensures consistent and reliable predictions regarding chemo-mechanical fracture in lithium-ion batteries.

	Eventually, it should be pointed out, numerical predictions on chemo-mechanical fracture, obtained by the proposed new phase-field interface Model-E2 with exponential interpolation of fracture energy, align with those based on the Model-S2 with stair-wise interpolation. This agreement is attributed to the achieved consistency between the phase-field integrated interface fracture energy and the sharp value defined in the classical cohesive zone model, which applies to concurrent bulk and interface fracture, as seen in \cref{eq:3D-energy-consistence}. 
	However, based on the aforementioned simulations of image-based 3D reconstructed polycrystalline cathode particles, the proposed CPF interface model with exponential interpolation of fracture property exhibits advantages over other models due to its high flexibility in handling structures containing complicated GBs topology, which is achieved through the auxiliary indicator variable solved from the Allen-Cahn equation.
	
	%%%%%%%%%%%%%%%%%%%%%%%%%%%%%%%%%%%%%%%%%%%%
	\section{Conclusions}
	\label{sec:conclusions}

	In this study, we proposed a new cohesive phase-field interface fracture model, the new model is on the basis of Euler-Lagrange equation of the variational problem and interface fracture energy check w.r.t that of the cohesive zone model. It utilizes an exponential function, which is constructed conveniently with the Allen-Cahn type auxiliary indicator, for the interpolation of fracture energy between the bulk phase and the interface within the structures consisting of complicated GBs topology; while the effective interface fracture energy $\tilde{G}_i$ is derived in such a way that the integrated phase-field fracture energy across the diffusive interface region remains consistent to that of the sharp interface fracture energy $G_i$ defined in the classical cohesive zone model.

	% To address inter-granular fracture at grain boundaries within the phase-field framework, we adopted a diffusive representation of the interface, incorporating an elaborately determined effective fracture toughness. In this paper, we extended the stairwise representation from existing literature to a general application, in particular, an exponential representation utilizing a predefined interface indicator, combined with the effective interface fracture toughness to ensure equivalence of crack surface energy to that of a sharp interface, is proposed.

	The proposed phase-field interface model is further extended to multi-dimensional formulation and seamlessly integrated into the chemo-mechanically coupled scenario in a thermodynamically consistent manner for the Lithium-ion battery cathode materials. The above multi-physically coupled cohesive phase-field fracture model is numerically implemented with finite-element method in an open-sourced platform MOOSE. 

	Benchmark examples of square plates with predefined notch and internal grain boundary under both mechanical and chemo-mechanical scenarios highlight following merits of the proposed model. Concurrent bulk (trans-granular) and interface (inter-granular) fracture behaviors can be simulated with the proposed unified model. Thanks to the constructed consistency between the integrated phase-field interface fracture energy and the sharp value defined in the cohesive zone model, numerical outcomes like crack evolution pattern, displacement and structural reaction force given by the proposed model consistently align with predictions from the CZM. In addition, numerical results, demonstrate insensitivity/independence with respect to length-scale parameters, i.e., the regularized thicknesses of phase-field fracture surface $b$ and interface $L$.

	Eventually, the coupled CPF model is applied to chemo-mechanical simulations of image-based reconstructed 3D NMC polycrystalline particles, which validates the high flexibility offered by our proposed phase-field interface Model-E2 in handling 3D polycrystalline structure with complicated GBs topology. The influences of sharp interface fracture energy and grain size on chemo-mechanical fracture behaviors are studied. It can be concluded from numerical simulations, the fracture energy of the GBs significantly affect the fracture modes in NMC polycrystalline particles; besides, particles with large grain size can yield both inter- and trans-granular cracks, while inter-granular fracture mode is dominantly exhibited for particles with small grain size.

	Looking forward, our outlooks encompass two potential areas for future investigation. Firstly, while our proposed model aligns with experimental findings and prior numerical studies by assuming GBs possess relatively weaker fracture energy compared to grains, there is merit in exploring scenarios where the GBs exhibits stronger fracture properties than the bulk materials. This perspective could offer valuable insights into computational phase-field fracture modeling. Secondly, the effect of electrolyte and its coupling with fracture propagation is ignored in the current manuscript. Depending on the (liquid/solid) state of employed electrolyte, the impacts of cracks on the chemical process can vary, e.g., cracks lead to an increase in electrochemically active surface areas and an improvement in charge-transfer kinetics due to the wetting effect \cite{Tanim2022AEM} in liquid electrolytes, while interface separation at electrode/solid electrolyte degrades and impede the kinetics \cite{Ruess2020JES,Stein2016JPS}. Such influence will be further incorporated into the current framework in our future studies.
    %the fatigue behaviors of electrodes under long-term operational scenarios also serve as a significant aspect that warrants further exploration in our forthcoming research endeavors.

	\clearpage
	\thispagestyle{empty}
	
	%%%%%%%%%%%%%%%%%%%%%%%%%%
	\section*{Acknowledgments}
	
	The authors Chen and Xu gratefully acknowledge the computing time granted on the Hessian High-Performance Computer "Lichtenberg" (Project-02017, Special-00007). This work has been (partially) funded by the German Research Foundation (DFG) under grant 460684687. The third author J. Y. Wu acknowledges the support from the National Natural Science Foundation of China (52125801). The authors would like to express deep gratitude to Dr. Jeffery M. Allen and Prof. Dr. Kandler Smith in National Renewable Energy Laboratory, Golden, USA, for their help with the generation of finite-element mesh of image-based reconstructed 3D NMC polycrystalline particles.
    
    %This work is conducted in part by the National Renewable Energy Laboratory, operated by Alliance for Sustainable Energy, LLC, for the U.S. Department of Energy (DOE) under Contract No. DE-AC36-08GO28308. Funding is provided by the U.S. DOE Office of Vehicle Technology Energy Storage Program, eXtreme Fast Charge and Cell Evaluation of Lithium-Ion Batteris (XCEL) Program, program manager Brian Cunningham. The views expressed in the article do not necessarily represent the views of the DOE or the U.S. Government. The U.S. Government retains and the publisher, by accepting the article for publication, acknowledges that the U.S. Government retains a nonexclusive, paid-up, irrevocable, worldwide license to publish or reproduce the published form of this work, or allow others to do so, for U.S. Government purposes.
	
	%%%%%%%%%%%%%%%%
	\appendix
	\gdef\thesection{\Alph{section}}
	\makeatletter
	\renewcommand\@seccntformat[1]{\appendixname\ \csname the#1\endcsname.\hspace{0.5em}}
	\makeatother
	
	\section{Analytical derivation of phase-field profile under stair-wise distribution of fracture energy}
	\label{appendix-analytical-stairwise}

		Follow previous work of \cite{Yoshioka2021CMAME,Zhou2022IJSS}, here an analytical derivation of phase-field profile $d(x)$ under following stair-wise $G(x)$, which is solved from Euler-Lagrange equation in \cref{eq:Euler-Lagrange-interface}, is briefly recalled,
		\begin{align}
			G(x) = 
			\begin{cases}
				G_i, &\text {when }|x|\leq L \\
				G_b, &\text {when }|x| > L
			\end{cases}.
		\end{align}
		A general solution under the stair-wise distribution of fracture energy can be:
		\begin{align}
			d(x) = 
			\begin{cases}
				1 - \epsilon_1 \cos \left( \dfrac{x}{b} \right) - \epsilon_2 \sin \left( \dfrac{x}{b} \right), \quad & \text{when } |x| \leq L \vspace{0.5mm} \\
				1 - \theta_1 \cos \left( \dfrac{x}{b} \right) - \theta_2 \sin \left( \dfrac{x}{b} \right), \quad & \text{when } L < |x| \leq b_0 \vspace{0.5mm} \\
				0, \quad & \text{when } |x| > b_0 \\
			\end{cases},
		\end{align}
		with $b_0$ denoting the location where phase-field variable reaches 0, i.e., $d(x = b_0) = 0$. Besides, $d(x = 0) = 1$ assuems the crack happens in the center of the domain and $d^{\prime}(x = b_0) = 0$ satisfies the irreversible evolution condition of phase-field variable. According to those conditions, we can eliminate two coefficients, i.e.,
		\begin{align}
			d(x) = 
			\begin{cases}
				1 - \epsilon \sin \left( \dfrac{x}{b} \right), \quad & \text{when } |x| \leq L \vspace{1mm} \\
				1 - \sin \left( \dfrac{x}{b} + \theta \right), \quad & \text{when } L < |x| \leq \left( \dfrac{\pi}{2} - \theta \right) b \vspace{1mm} \\
				0, \quad & \text{when } |x| > \left( \dfrac{\pi}{2} - \theta \right) \\
			\end{cases},
		\end{align}
		with $\epsilon_1 = 0$, $\epsilon_2 = \epsilon$, $\theta_1 = \sin \theta$, $\theta_2 = \cos \theta$ and $b_0 =  \left( \dfrac{\pi}{2} - \theta \right) b$ obtained from direct mathematical derivations.

		Due to the discontinuity of $G(x)$ at $x = L$, Weierstrass-Erdmann corner conditions \cite{gelfand2000calculus} are further supplemented, 
		\begin{align}
			\begin{cases}
				d\left( x = L^{-} \right) = d \left( x = L^{+} \right) \\
				G_i \; d^{\prime} \left( x = L^{-} \right) = G_b \; d^{\prime} \left( x = L^{+} \right)
			\end{cases}
			\quad \Longrightarrow \qquad
			\begin{cases}
				\epsilon = 1/ \sqrt{\sin^2 \left( \dfrac{L}{b} \right) + \left( \dfrac{G_i}{G_b} \right) ^2 \cdot \cos^2 \left( \dfrac{L}{b} \right) } \vspace{2mm} \\
				\theta = \tan ^{-1} \left[ \dfrac{G_b}{G_i} \tan \left( \dfrac{L}{b} \right) \right] - \dfrac{L}{b}
			\end{cases},
		\end{align}
		so that $\epsilon$ and $\theta$ can be further solved and expressed as above.

		Note phase-field profile $d(x)$ in the interface Model-S2 in \cref{eq:profile-Model-S2} can also be derived in a similar fashion, by replacing $G_i$ with the effective value $\tilde{G}_i$ in the above. 

	\section{Formulations of cohesive zone model for chemo-mechanical fracture}
	\label{appendix-chemo-mechanical-czm}
	
		In this section, the chemo-mechanical cohesive zone model for simulating fracture in Lithium-ion batteries is briefly recalled. The governing equations and corresponding constitutive laws of the bulk ($\varOmega$) for chemical and mechanical sub-problems are listed as following.
		\begin{subequations}\label{eq:governing-equation-bulk}
			\begin{align}
				&\begin{cases}
					\nabla \cdot \boldsymbol{J} + \dot{c} = 0 \\
					\mu = R T \ln \dfrac{\tilde{c}}{1-\tilde{c}} - \bfsigma : \boldsymbol{\Omega} \\
					\boldsymbol{J} = - c(1-\tilde{c})/(RT) \boldsymbol{D} \cdot \nabla \mu \\
				\end{cases}\\
				&\begin{cases}
					\nabla \cdot \bfsigma = \boldsymbol{0} \\
					\bfsigma = \mathbb{E}_0 : (\bfepsilon - \bfepsilon_c)
				\end{cases}
			\end{align}
		\end{subequations}

		With particular interest and focus on fracture behaviors, here the diffusion across the GB/interface ($\varGamma$) is assumed to be coherent and continuous for simplicity, while the crack evolution on the interface is captured with cohesive zone model as following,
		\begin{align}\label{eq:governing-equation-interface}
			\int_{\varGamma} \left( 
					\boldsymbol{t}^{+} \cdot \delta \boldsymbol{u}^{+} + 
					\boldsymbol{t}^{-} \cdot \delta \boldsymbol{u}^{-} 
					\right) \; \mathrm{d} A = 
			\int_{\varGamma}  
					\boldsymbol{t} \cdot \delta [\![ \boldsymbol{u} ]\!] 
					\; \mathrm{d} A
		\end{align} 
		which can be obtained by applying divergence theorem to strong form in \cref{eq:governing-equation-bulk}, here $\boldsymbol{t}^{+} = - \boldsymbol{t}^{-} = \boldsymbol{t}$ satisfies the mechanical equilibrium at the interface, and $\boldsymbol{t}$ denotes the traction force. In the CZM, the displacement jump is adopted to characterize the fracture evolution, i.e., total jump in global coordinate system is expressed as $[\![ \boldsymbol{u} ]\!] = \boldsymbol{u}^{+} - \boldsymbol{u}^{-}$, so that the jump in interface coordinate system $[\![ \tilde{\boldsymbol{u}} ]\!] = \tilde{\boldsymbol{R}} [\![ \boldsymbol{u} ]\!]$ can be obtained with rotation matrix $\tilde{\boldsymbol{R}}$.

		The local traction force at the interface $\tilde{\boldsymbol{t}}$ can be determined from the cohesive zone constitutive model, i.e., traction-separation law (TSL), w.r.t. to the local displacement jump $[\![ \tilde{\boldsymbol{u}} ]\!]$, then traction force in global coordinate system $\boldsymbol{t}$ (\cref{eq:governing-equation-interface}) can be obtained. In the modeling, bilinear softening law \cite{camanho2002mixed} is adopted to characterize tension-dominant fracture behaviors. For further details, readers are referred to \cite{Shahed2021JMPS,Bai2021IJSS}, including more complicated contexts such as mixed-modes fracture and large mechanical deformation setup.

	\bibliographystyle{elsarticle-num}
	\biboptions{numbers,sort&compress}
	
	\bibliography{refer}
	
\end{document}